\documentclass[iop]{emulateapj}

\usepackage{graphicx}
\usepackage{array}
\usepackage{dcolumn}
\newcolumntype{d}{D{.}{.}{2}}
\newcommand{\NA}{\multicolumn{1}{c}{\mbox{NA}}}

\usepackage{color}
\usepackage{ulem}
\usepackage[english]{babel} 
\usepackage{amssymb,amsmath}

\newcommand{\dxdy}[2]{{\frac{\partial{#1}}{\partial{#2}}}}

\newcommand{\DxDy}

\def\rhobar{{\bar\rho}}

\def\Tbar{{\overline T}}

\def\gbar{{\bar g}}

\def\nubar{{\bar\nu}}

\def\etabar{{\bar\eta}}

\def\rhat{{\hat{\bf r}}}

\def\omegavec{{\bf \Omega}}
\def\grad{{\bf \nabla}}
\def\div{{\bf \nabla} \cdot}
\def\curl{{\bf \nabla} \times}
\def\vvec{{\bf v}}

\def\Bvec{{\bf B}}
\def\Evec{{\bf E}}
\def\Jvec{{\bf J}}

\def\eten{e_{ij}}
\def\dten{\delta_{ij}}

\newcommand{\myemail}{tami@lpl.arizona.edu}

\shorttitle{Magnetism in Hot Jupiters}
\shortauthors{T.M. Rogers and T.D. Komacek}

\begin{document}

\title{Magnetic Effects in Hot Jupiter Atmospheres}

\author{T.M. Rogers and T.D. Komacek} \affil{Department of Planetary Sciences,
 University of Arizona, Tucson, AZ, 85721}
\email{\myemail}
\begin{abstract}
We present magnetohydrodynamic (MHD) simulations of the atmospheres of hot Jupiters ranging in temperature from 1100-1800K. Magnetic effects are negligible in atmospheres with temperatures $\lesssim$ 1400K. At higher temperatures winds are variable
and in many cases, mean equatorial flows can become westward, opposite to their hydrodynamic counterparts. Ohmic dissipation peaks at temperatures $\sim$1500-1600K, depending on field strength, with maximum values $\sim 10^{18}$W at 10bar, substantially lower than previous estimates. Based on the limited parameter study done, this value can not be increased substantially with increasing winds, higher temperatures, higher field strengths, different boundary conditions or lower diffusivities. Although not resolved in these simulations there is modest evidence that a magnetic buoyancy instability may proceed in hot atmospheres. 
\end{abstract}


\section{Introduction}
\indent Exoplanet research is an expeditiously changing field, driven by the discovery of more than 1,450 exoplanets to date. As many of these exoplanets are discovered either by the radial velocity or transit technique, there is a natural bias towards discovering planets that both have large radii and are close to their host star. These close-in extrasolar giant planets (EGPs), or ``hot Jupiters'', are nominally assumed to be tidally locked to their host star and have strong atmospheric flows driven by their high irradiation, $10^3-10^6$ times the flux that Jupiter receives. This results in the unique atmospheric dynamics of these hot Jupiters, with equatorial superrotation and strong day-night flow predicted to occur from purely hydrodynamical numerical simulations \citep{cho03,show02,coop05,iandd08,show09,rm10,lewis10,tcho10,heng11}. These simulations also predict eastward displacement of the hot spot from the substellar point, first predicted by \cite{show02} and confirmed through observations of HD 189733b by \cite{knutson07,knutson12}. However, the extremely high temperatures present in many hot Jupiter atmospheres point to high levels of ionization, which could lead to magnetic effects, which we discuss in this work. \\
\indent The combination of radial velocity data, transit light curves, and spectrometry have enabled observations to place strong constraints on the mass, radius, composition, temperature, and wind structure of hot Jupiters, leading them to be the most well-characterized planets outside our Solar System. However, there are still many unexplained observations, such as: the extreme variability of the dayside-nightside circulation efficiency of these planets \citep{cowan11}, and the larger than expected radii of many hot Jupiters \citep{bod01,bod03,guillot02,ba03,laughlin05}. The circulation efficiency of EGP is determined by combining information from infrared and visible phase curves, with some planets (e.g. HD 209458b, HD 189733b) having very efficient circulation from dayside to nightside and hence small ($ < 500$ K) zonal (east-west) temperature variations \citep{cowan07,knutson07}, and some (e.g. Ups And b, WASP-12b) having weak circulation and large day-night temperature variations \citep{crossfield10,cowan12}. Using a statistical approach of 24 transiting exoplanets with light curves \cite{cowan11} conclude that hot Jupiters show a variety of recirculation efficiencies, with no clear trend with temperature. However, they do note that the hottest planets in their sample have the lowest recirculation efficiencies, with a fairly sharp transition at $T_{\mathrm{eff}} \approx 2400$K. Reduced circulation efficiencies might be expected at high temperatures where magnetic fields are able to reduce wind speeds. \\
\indent The enhanced radii of many hot Jupiters is likely solved by invoking an additional heat source in the giant planet interior to counteract their gravitational contraction. A variety of heating mechanisms have been proposed, e.g.: tidal heating \citep{bod01,bod03,jackson08}, deposition of energy from waves \citep{guillot02}, forced turbulence \citep{youdin10}, and most recently, Ohmic dissipation \citep{bat10,perna10b,laughlin11}. The exact amount of heating necessary to inflate these hot Jupiters is uncertain, but is approximately $10^{-2}-10^{-6}$ of the stellar incident flux, depending on the composition of the planet and the depth at which this energy is deposited \citep{bod01,show02}. Using 90 well-characterized EGP \cite{laughlin11} found that the radius anomaly (the difference between the observed radius and predicted radius from evolutionary models, $\Delta R = R_{\mathrm{obs}} - R_{\mathrm{pred}}$) is positively correlated with the effective temperature of the planet. This points toward Ohmic dissipation as a viable mechanism, which is supported by recent models \citep{bat10,perna10b,wu13}. Though other heating mechanisms may explain a portion of the observed radii \citep{ba03}, \cite{bat11} claim that only Ohmic dissipation coupled with metallicity variations can explain the entire spectrum of hot Jupiter radii observations. 

Coupling the effects of magnetic drag (which reduces wind speeds) and Ohmic dissipation, \cite{menou12} showed that Ohmic dissipation has a peak temperature beyond which it decays due to the strong influence of the Lorentz force slowing wind speeds past a critical equilibrium temperature. For a 10G field, \cite{menou12} finds Ohmic dissipation peaks at $\sim 1600$K, remarkably close to the peak seen in the \cite{laughlin11} data for radius anomaly, adding further confidence that Ohmic dissipation is the explanation for inflated hot Jupiter radii. However, more recently the viability of Ohmic dissipation as the heating mechanism has been questioned. Adding the scaling laws of \cite{menou12} to a one-dimensional planetary evolution model, \cite{huang12} found Ohmic dissipation to be too weak to explain the radii of hot Jupiters with masses $\gtrsim 0.4 M_{\mathrm{Jup}}$. Using a similar radiative-convective model but including the effects of nightside cooling (making the model 1+1D), \cite{sb13} found that Ohmic dissipation cannot by itself produce a runaway in planet radius as predicted by \cite{bat11}. Additionally, the recent General Circulation Models (GCMs) of HD 209458b and HD 189733b by \cite{rm13}, using a kinematic model for wind drag and a parametrization of Ohmic dissipation, could not explain the inflated radii of those planets for dipole magnetic field strengths $\lesssim 10$G without assuming a large interior heat flux and extrapolating using scalings from \cite{wu13}. 

In the first self-consistent MHD simulations of a hot Jupiter atmosphere which included Ohmic dissipation \cite{rogers14} showed that Ohmic dissipation fell short of explaining the inflated radius of HD209458b by nearly two orders of magnitude. This is in stark contrast with the results of \cite{bat10,bat11} and \cite{perna10b}, who find that the Ohmic dissipation at depth can explain, and even over-predict, the radii of the population of hot Jupiters. However, the models of \cite{bat10} and \cite{bat11} assumed that a fixed amount of the incident stellar flux went into Ohmic dissipation, and \cite{perna10b} did not include feedback from magnetic drag in their GCMs. Due to the variety of methodologies used to explore Ohmic dissipation and the extreme sensitivity of each model to parameters that vary from planet to planet, there are large discrepancies in the literature on magnetic effects in hot Jupiter atmospheres and interiors. 

Currently, many of the unexplained observations of hot EGP atmospheres have been attributed to planetary magnetic fields. However, only two self-consistent MHD models of the interaction of planetary magnetic fields with atmospheric dynamics has been conducted \citep{bat13,rogers14} and only one \citep{rogers14} includes Ohmic dissipation. Such models are necessary if these theories are to be made robust. Here we extend the work of \cite{rogers14} to cover a range of atmospheric temperatures (and hence, magnetic diffusivities) to understand the breadth of possible magnetic effects in hot Jupiter atmospheres. 
\\
\section{Numerical Methods}
We use a three-dimensional (3D), magnetohydrodynamic (MHD) model in the anelastic approximation. The model is based on the Glatzmaier dynamo code \citep{glatz84,glatz85}, but has substantial differences in the actual equations solved, the discretization and the implementation. A description of the axisymmetric version of this code can be found in \cite{rogers11}. The model solves the following equations:  
\begin{eqnarray}
\div \rhobar \vvec &=& 0
\\
\div \Bvec &=& 0
\\
\rho &=& \left[
\left( {\overline {\partial \rho \over \partial T}} \right )_{p} T +
\left( {\overline {\partial \rho \over \partial p}} \right )_{T} p
\right]
\\
\rhobar {\partial \vvec \over \partial t} &=&
-\div ( \rhobar \vvec \vvec )
- \grad p \
- \rho \gbar \rhat
\\ \nonumber
&& + 2 \rhobar \vvec \times \omegavec
+ \div (2 \rhobar \nubar
(\eten - {1 \over 3} ( \div \vvec ) \dten)) \\
&& +{1 \over \mu_0} ( \curl \Bvec ) \times \Bvec \nonumber
\\
{\partial \Bvec \over \partial t} &=&
\curl ( \vvec \times \Bvec )
- \curl (\overline{\eta} \curl \Bvec)
\\
\dxdy{T}{t}&+&(\vec{v}\cdot\vec{\nabla}){T}= \\
&& -v_{r}(\dxdy{\overline{T}}{r}-(\gamma-1)\overline{T}h_{\rho})+(\gamma-1)Th_{\rho}v_{r} \nonumber
\\ \nonumber
&& +\gamma\overline{\kappa}[\nabla^{2}T+(h_{\rho}+h_{\kappa})\dxdy{T}{r}] + \frac{T_{eq}-T}{\tau_{rad}}+\frac{\overline{\eta}}{\mu_{o}\overline{\rho} c_{p}}|\nabla \times {\bf B}|^{2}
\end{eqnarray}
Equation (1) represents the continuity equation in the anelastic approximation \citep{ogura62,gough69,glatz84,rg05}, where $\rhobar$ is the reference state density and $\vvec$ is the gas velocity. The anelastic approximation was developed for dealing with convection and winds in Earth's atmosphere and therefore, does not fully capture the large atmospheric wind velocities likely on hot Jupiters. While not ideal, the progress obtained by performing full MHD models (as opposed to the kinematic models previously executed) vastly outweighs the inaccuracy associated with slightly lower wind speeds.\footnote{Note that this form of the anelastic equations does not conserve energy for linear perturbations \citep{durran89}. In order to test whether this had an effect in the flows described here we implemented the fix suggested by \cite{brown12} in our hydrodynamic models.  After 1200 P$_{rot}$ we find the flows are qualitatively virtually identical.  Quantitatively, the energy conserving formalism suggested in \cite{brown12} leads to maximum flow speeds which are slightly larger (5\%).}  Equation (2) represents the conservation of magnetic flux and Equation (3) is the equation of state, which we take here to be an ideal gas. The momentum equation, including Coriolis and Lorentz forces, is represented by Equation (4) where $p$ represents pressure, $\omegavec$ is the angular velocity of the rotating frame of reference, $\eten$ is the rate of strain tensor, $\nubar$ is the viscous diffusivity and $\Bvec$ is the magnetic field. 

The magnetic induction equation is represented by Equation (5), where $\etabar$ is the magnetic diffusivity (see below for treatment of the magnetic diffusivity).  The energy Equation (6) is written here as a temperature equation with $\Tbar$ the reference state temperature, $\kappa$ the thermal diffusivity and $h_{\rho}$, $h_{\kappa}$ representing inverse density and thermal diffusivity scale heights, respectively. We note that the thermal and viscous diffusivities in Equations (4) and (6) can be functions of radius. The first term on the right hand side (RHS) of Equation (6) represents the super- or subadiabaticity of the region and therefore, allows us to treat both convective and stably stratified regions, which is unnecessary in this work but may be necessary in the future. The fourth term on the rhs of Equation (6) represents the Newtonian radiative forcing we impose to mimic the differential day-night heating imposed by the host star, where $\tau_{rad}$ is a function that varies between $10^{4}$s at the outer most layers and $10^{6}$s at the lowest layers. The equilibrium temperature is given by the simple relation 
\begin{equation}
T_{eq}(r,\theta,\phi)=\Tbar(r)+\Delta T_{eq} \cos\theta\cos\phi \mathrm{,}
\end{equation}
where $\theta$, $\phi$ are latitude and longitude and $\Delta T_{eq} $ is the specified day-night temperature variation. $\Tbar(r)$ is the reference state temperature of the model (see Figure 1). It, as well as the other thermodynamic reference state variables denoted with an overbar above, are taken from the planetary evolution model of \cite{iro05}.
\begin{figure}
\centering
\includegraphics[width=3.5in]{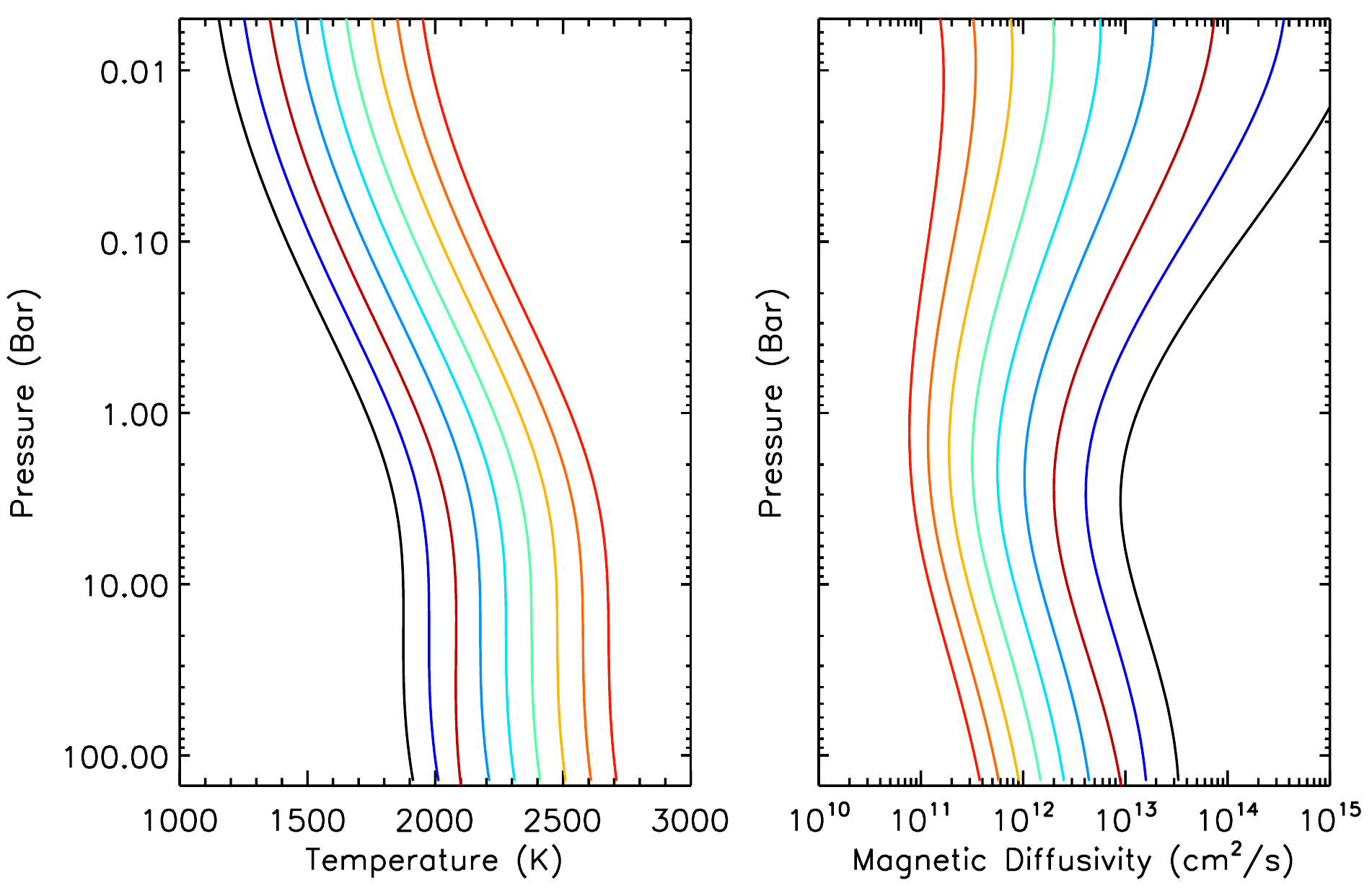}
\caption{Temperature-Pressure profiles for the models run. Left figure: models M1-M9, from left to right. Right: Magnetic Diffusivity-Pressure profiles from right to left models M1-M9.} 
\label{fig:tpeta}
\end{figure}
The last term on the RHS of (6) represents Ohmic heating. We calculate the magnetic diffusivity, $\etabar$, given our reference state density and temperature profile using the method from \cite{rm13}:
\begin{equation}
\etabar = 230 \frac{\sqrt{\overline{T}}}{\chi_e} \mathrm{cm}^2 \mathrm{s}^{-1} \mathrm{,}
\end{equation}
where $\chi_e$ is the ionization fraction calculated from the Saha equation. We use a form of the Saha equation taking into account all elements from hydrogen to nickel, using elemental abundances from \cite{Lodders10}, and calculate
\begin{equation}
\frac{\chi_{e,i}^2}{1 - \chi_{e,i}^2} = n_i^{-1} \left(\frac{2\pi m_e}{h^2}\right)^{3/2} (kT)^{3/2} \mathrm{exp}(-\epsilon_i / kT) 
\end{equation}
for each element $i$, where $\epsilon_i$ is its ionization potential, $m_e$ the mass of the electron, $h$ Planck's constant, $k$ the Boltzmann constant, and the number density $n_i$
\begin{equation}
 n_i = n\left(\frac{a_i}{a_H}\right) \mathrm{,}
\end{equation}
where $a_i/a_H$ is the abundance of each element normalized to that of hydrogen and $n = \rho/\mu$, with $\mu$ the solar mean molecular weight from \cite{Lodders10}. The total ionization fraction is found through summation:
\begin{equation}
\chi_e = \sum\limits_{i=1}^{28} \frac{n_i}{n} \chi_{e,i} \mathrm{.}
\end{equation}
This total ionization fraction is inserted into Equation (8) to compute $\etabar$ (see Figure 1b). We assume solar metallicity for all simulations, as we do not calculate evolutionary models and an enhanced metallicity is similar in effect to a hotter temperature. 
 
Equations (1)-(6) are solved using the spherical harmonic poloidal-toroidal decomposition for the divergence-free mass flux and magnetic field employed for decades in 3D MHD simulations. The radial dimension is discretized using a fourth-order finite difference scheme. Timestepping is a combination of the explicit Adams-Bashforth method for the nonlinear terms and an implicit Crank-Nicolson method for the linear terms. 
The velocity boundary conditions are stress-free and impermeable (although see model M7cbc*), the temperature boundary conditions are zero perturbation at top and bottom. The dipole magnetic field is matched to a potential field at the top boundary and is set to a constant (the magnetic field strength for a given model) at the bottom boundary. The current is set to zero at both boundaries. The model is parallelized using Message Passing Interface (MPI) and simulations have resolution 128 ($N_{\phi}$) x 64 ($N_{\theta}$) x 100 ($N_{r}$). All models have been run for at least 1500 rotation periods, which generally requires $\sim$5000 processor hours per simulation (although this number varies substantially for the various models).

\subsection{Setup}
\begin{table*}
\renewcommand{\tabcolsep}{5pt}
\renewcommand{\baselinestretch}{1}
\small
 \centering
 \begin{tabular}{ldddddddddddd}
 Model &\textrm{T}& \textrm{B}_{o} &\nu & \kappa & \eta & \Delta T_{eq} & \tau_{mag} & B_{\phi \mathrm{m}} & \mathrm{OD} & \mathrm{WS} & \mathrm{EFF} \\
 \hline
 M1 & 1100 & 0 & 5. & 5. & \NA & 800 &\NA &\NA&\NA&2.8 & \NA\\
 M1b1 & 1100 & 10 & 5. & 5. & 1260. &800&$2.06$\mathrm{E}$9$&0.78&0.12&2.8 & 3.6 \\
 M1b2 & 1100 & 30 & 5. & 5. & 1260. & 800&$2.37$\mathrm{E}$8$&2.3&1.0& 2.8 & 30.3\\
 
 M2 & 1200 & 0 & 5. & 5. & \NA & 800 &\NA&\NA&\NA&2.9 & \NA\\
 M2b1 & 1200 & 10 & 5. & 5. & 580. &800 &$1.03$\mathrm{E}$8$&2.2&0.33&2.9 & 10.0\\
 M2b2 & 1200 & 30 & 5. & 5. & 580. & 800 &$1.4$\mathrm{E}$7$&6.0&3.0&2.9 & 90.9\\
 
 M3  & 1300 & 0 & 5. & 5. & \NA & 800 &\NA&\NA&\NA&3.0 & \NA \\
 M3b1 & 1300 & 10 & 5. & 5. & 360. &800 &$3.6$\mathrm{E}$6$&7.2&0.63&3.0 & 19.1\\
 M3b2 & 1300 & 30 & 5. & 5. & 360. & 800 &$4.66$\mathrm{E}$5$&20.1&4.8&3.0 & 145.5\\
 
 M4 & 1400 & 0 & 5. & 5. & \NA & 800 &\NA &\NA&\NA&3.1 & \NA\\
 M4b1 & 1400 & 10 & 5. & 5. & 150. &800&$2.9$\mathrm{E}$5$&18.6&3.7&3.1 & 112.1\\
 M4b2 & 1400 & 30 & 5. & 5. & 150. & 800&$5.0$\mathrm{E}$4$&44.6&21.0& 3.5& 636.4\\
 
 M5 & 1500 & 0 & 5. & 5. & \NA & 800 &\NA &\NA&\NA&3.1 & \NA\\
 M5b1 & 1500 & 10 & 5. & 5. & 84. &800 &$2.1$\mathrm{E}$4$&48.5&13.&3.3 & 393.9\\
 M5b2 & 1500 & 30 & 5. & 5. & 84. & 800 &$7.5$\mathrm{E}$3$&81.3&28.&2.9 & 848.5\\
 M5b3 & 1500 & 3 & 5. & 5. & 84. & 800 &$2.0$\mathrm{E}$5$&15.8&1.3&3.1& 3.4\\ 
 
 M6 & 1600 & 0 & 5. & 5. & \NA & 800 &\NA &\NA&\NA&3.2 & \NA\\
 M6b1 & 1600 & 10 & 5. & 5. & 48. &800&$2.1$\mathrm{E}$3$&111.&26.&2.6 & 787.9\\
 M6b2 & 1600 & 30 & 5. & 5. & 48. & 800&$1.26$\mathrm{E}$3$&144.&25.&2.2 & 757.6\\
 M6b3 & 1600 & 3 & 5. & 5. & 48. & 800 &$1.05$\mathrm{E}$4$&49.8&5.2&3.2 & 157.6\\

 M7 & 1700 & 0 & 5. & 5. & \NA & 800 &\NA&\NA&\NA&3.2 & \NA\\
 M7b1 & 1700 & 10 & 5. & 5. & 29. &800 &263.8&234.5&20.&2.3 & 606.1\\
 M7b2 & 1700 & 30 & 5. & 5. & 29. & 800 &317.&213.9&22.&2.0 & 666.7\\
 M7b3 & 1700 & 3 & 5. & 5. & 29. & 800 &995.8&120.7&13.&3.0 & 393.9\\

 M7v1 & 1700 & 0 & 0.5 & 0.5 & \NA &800 &\NA&\NA&\NA&2.7 & \NA\\
 M7v1b1 & 1700 & 10 & 0.5 & 0.5 & 29. & 800&320.&212.9&30.6&2.2 & 927.3\\
 
 M7v2 & 1700 & 0 & 0.05 & 0.05 & \NA &800 &\NA&\NA&\NA&2.9 & \NA\\
 M7v2b1 & 1700 & 10 & 0.05 & 0.05 & 29. & 800&404.8&189.3&36.&2.0 & 1090.9\\
  
 M7v3 & 1700 & 0 & 0.02 & 0.02 & \NA &800 &\NA&\NA&\NA&2.9 & \NA\\
 M7v3b1 & 1700 & 10 & 0.02 & 0.02 & 29. & 800&417.&186.5&38.5&2.5 & 1166.7\\

 M7f1 & 1700 & 0 & 5. & 5. & \NA&1200 &\NA&\NA&\NA&3.6 & \NA\\
 M7f1b1 & 1700 & 10 & 5. & 5. & 29. & 1200 &228.4&239.& 29.&2.6 & 878.8\\
 M7f1b2 & 1700 & 30 & 5. & 5. & 29. & 1200 &253.9&252.& 30.&2.4 & 909.1\\

 M7cbc & 1700 & 0 & 5. & 5. & \NA&800 &\NA&\NA&\NA&2.3 & \NA\\
 M7cbcb1 & 1700 & 10 & 5. & 5. & 29.& 800&335.3&208.&17.&1.5 & 515.2\\
 
 M8 & 1800 & 0 & 5. & 5. & \NA&800 &\NA&\NA&\NA&2.9 & \NA\\
 M8b2 & 1800 & 10 & 5. & 5. & 18. & 800&215.7&208.9&12.&2.3 & 363.6\\
 M8b3 & 1800 & 3 & 5. & 5. & 18. & 800 &195.7 &199.0&23.&3.0 & 670.0\\
 
 M9 & 1900 & 0 & 5. & 5. & \NA &800 &\NA&\NA&\NA&2.9 & \NA\\
 M9b3 & 1900 & 3 & 5. & 5. & 12. & 800&69.1&278.2&17.5&3.2 & 530.3 \\
 %\hline
  \end{tabular}
  \normalsize
  \caption{Model parameters. T is the temperature of the reference state model, $\Tbar$, at 2 mbar in K, see Figure~\ref{fig:tpeta}. B$_{o}$ is the model field strength, ``0'' indicates a hydrodynamic model, other values are in Gauss and represent the radial field at the pole, at the bottom of the domain. Viscous ($\nu$), thermal ($\kappa$), and magnetic ($\eta$) diffusivities are in units of $ 10^{10}$ cm$^{2}$/s and are values at 10 Bar. $\Delta T_{eq}$ is the day-night forcing described in Equation (7). The magnetic timescale is in seconds and is calculated using the maximum toroidal field strength, $B_{\phi \mathrm{m}}$ at 1500 P$_{\mathrm{rot}}$ and the magnetic diffusivity at 0.5Bar (the region where the toroidal field strength peaks). OD is the Ohmic dissipation integrated below 10 Bar in $10^{18}$ Watts and WS is the peak wind speed at 1500 P$_{\mathrm {rot}}$ in km/s.  EFF is the efficiency of conversion from incident stellar flux to Ohmic dissipation in parts per million.}
\end{table*}
We ran a suite of models varying the reference state temperature, resulting in temperatures which vary 
at 2 mBar from 1100-1900K. As our fiducial model we used the model of HD209458b from \cite{iro05}. For our hotter models we add 100-900K at every pressure level. When we refer to temperature in the following sections we will be referring to this reference state temperature ($\Tbar$) at 2 mBar (the lowest pressures). The highest pressure level in our model (greatest depth) is 200 bar. This represents approximately 15\% of the planetary radius.  The temperature-pressure profiles for our various models are shown in Figure~\ref{fig:tpeta}a. We run the purely hydrodynamic models for 200 rotation periods (P$_{\mathrm{rot}}$) before adding a magnetic field, we then continue both hydrodynamic and MHD models for an additional 1300 rotation periods. Magnetic field strengths between 3-30G were were run, although not every field strength was run at every temperature (see Table 1). The quoted magnetic field strength is that of the radial field at the pole and at the base of the simulated atmosphere, in all cases the magnetic field has an initial radial profile which varies with radius as a dipole (falls off as $r^{-3}$). Using the prescription described in Section 2 we calculate the magnetic diffusivity as a function of height for each of those models which is shown in Figure~\ref{fig:tpeta}b. The day-night temperature forcing, $\Delta T_{eq}$, is set to 800K, however, 1200K was also tested (M7f1*, see Table 1). The rotation rate is taken to be three days and is not varied. The radiative timescale for the Newtonian forcing is fixed to that of the base \cite{iro05} model in all runs, in order to better ascertain how the MHD modifies day-night temperature differences. The fiducial thermal and viscous diffusivities are taken to be $5\times 10^{10}$ cm$^{2}$/s and are fixed with radius. However, Models M7v1*, M7v2* and M7v3* have diffusivities which vary as a function of depth with values at higher pressures orders of magnitude lower than those at lower pressures (see Table 1 for values). 
\section{Results}
\begin{figure*}
 \centering
 \includegraphics[width=6in]{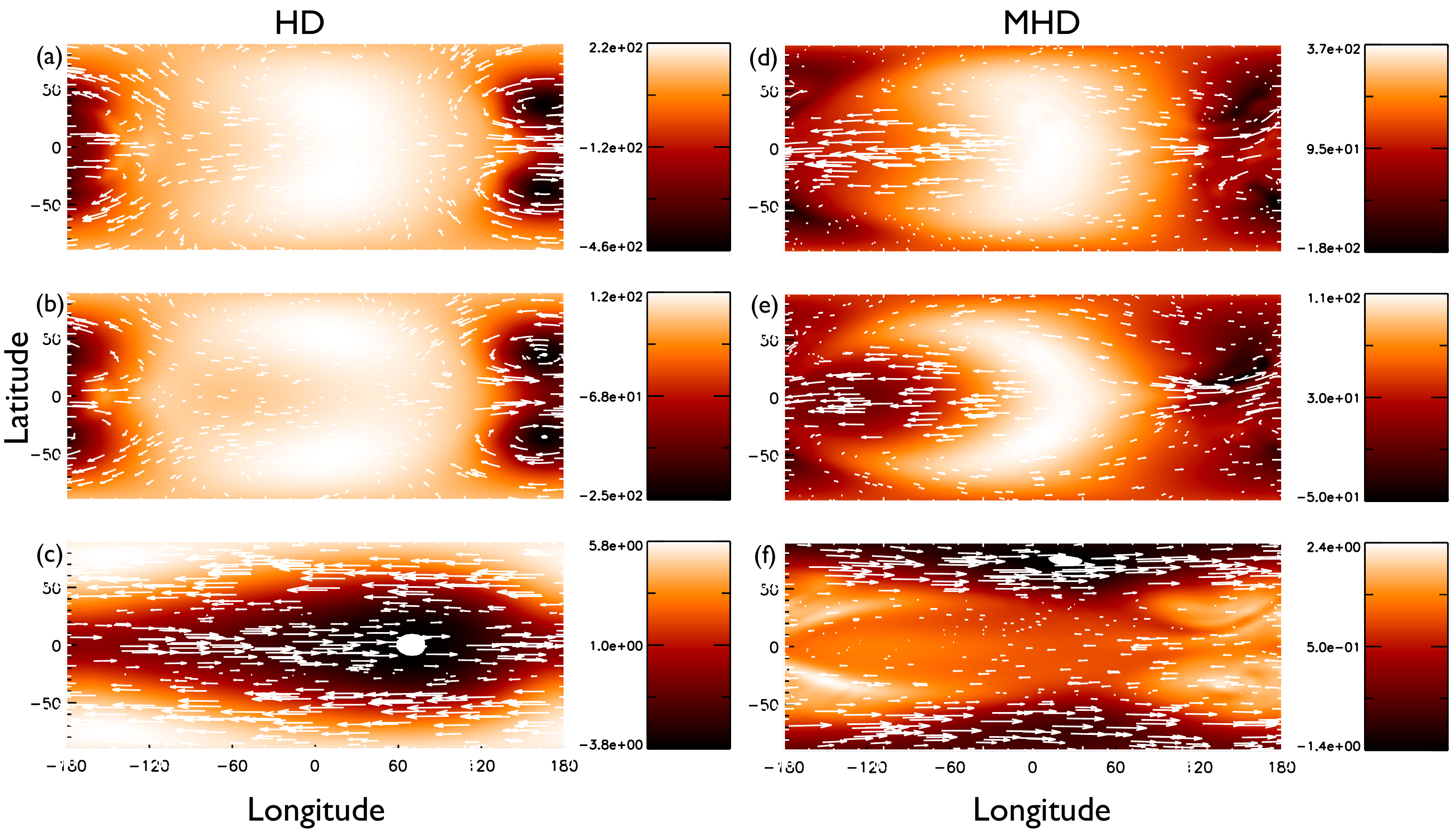}
 \caption{Temperature perturbation (K), shown in color, and winds, shown with arrows as a function of longitude and latitude, for models M7 (a,b,c) and M7b2 (d,e,f) at 10mBar (a,d), 70mBar (b,e) and 10Bar (c,f).}
 \label{fig:comp-tempwinds-700}
\end{figure*}
In general, the models with 
T $\lesssim$ 1400K show virtually no magnetic effects and we include
them for completeness, but will not discuss them in any detail.
Figure~\ref{fig:comp-tempwinds-700} shows temperature-wind maps
for one of our models (M7) showing both the hydrodynamic version (a,b,c) as
well as the magnetic version (M7b2,d,e,f) at three different heights and at the same times (see caption for details). The hydrodynamic models show behavior similar to previous models of hot Jupiter winds: a chevron like pattern which leads to predominantly eastward flow at the equator. Weak westward flows are seen high in the atmosphere and at some longitudes, but eastward flow is dominant at the equator. The hot spot is displaced eastward of the sub-stellar point (which is at 0$^{\circ}$ longitude). Meridonal flow is poleward westward of the hot spot and equatorward eastward of the hotspot. There are also a couple discrepancies compared with previous models. First, despite predominantly eastward flow near the equator (and eastward flow on zonal average) we do see some longitudes which show westward flow. Second, the temperature structure at depth with cool equator and hot poles (Figure~\ref{fig:comp-tempwinds-700}c) is opposite to that previously found. Both these may be due to the use of explicit thermal diffusivities which changes slightly the overall thermal forcing.

Comparing the hydrodynamic flows to the magnetic models we note several
differences: 1) the magnetic model is hotter overall, 2) the magnetic model shows smaller day-night temperature differences, 3) the magnetic model has predominantly westward flow, 4) the magnetic model has weaker meridional flow and 5) the magnetic model is asymmetric about the equator. We will address many of these differences in the following sections. 
\subsection{Magnetic Field Evolution}
\begin{figure*}
 \centering
 \includegraphics[width=6in]{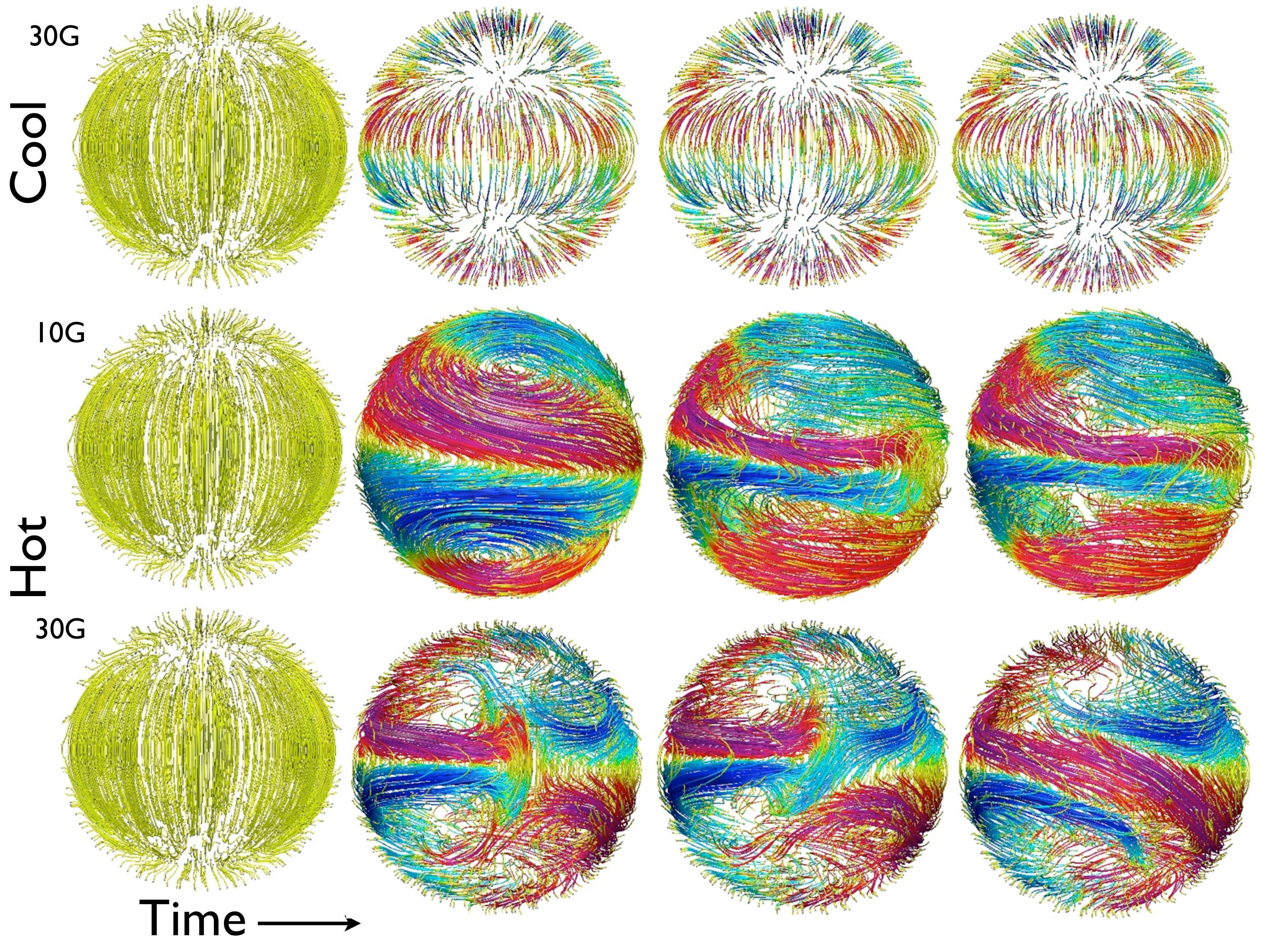}
 \caption{Magnetic field evolution. The viewpoint is looking onto the nightside of the planet. Top row shows field lines for M2b2, with color representing the toroidal field magnitude, with red/magenta positive (with maximum of 5G), blue negative (with minimum of -5G) and yellow representing values between $\pm$1G. Middle and bottom rows show field lines for M7f1b1 and M7f1b2, respectively. Again, color represents toroidal field strength, with red/magenta positive with maximum of 260G, blue negative with minimum -260G and yellow representing field strengths in the range $\pm$20G. Times are different for each model and are meant only to give a qualitative picture of magnetic field evolution.}
 \label{fig:fieldevolution-back}
\end{figure*}
For cooler models, the imposed field is virtually unaffected by the flow and there is no field evolution, virtually no induced toroidal field or current, hence little Lorentz force and little Ohmic dissipation. This can be seen in the top row of Figure~\ref{fig:fieldevolution-back}, which shows the evolution of magnetic field in models M2b2. There we see that the initially imposed dipole field is barely distorted. For hot planets, in which the field is sufficiently tied to the flow, field lines are swept toward the nightside of the planet by winds flowing away from the substellar point (both eastward and westward). Because eastward flow is favored, toroidal field induction is predominantly positive in the Northern Hemisphere (NH) and negative in the Southern Hemisphere (SH) near the equator. Since the winds are dragged by the Lorentz force and field is swept both eastward and westward, the field is not continually wrapped around the circumference of the planet. Rather, field of opposite signs meets on the nightside of the planet, causing toroidal field strengths and currents to peak there. The resulting large gradients in the toroidal field cause strong variation in the Lorentz force and unsteady flow and field, particularly for stronger fields and/or higher temperatures (note reversals in bottom row of Figure~\ref{fig:fieldevolution-back}). This leads to variability and asymmetry in the zonal flows and hence, variability in hot-spot displacement, see Section 3.4.
\begin{figure}
 \centering
 \includegraphics[width=3.5in]{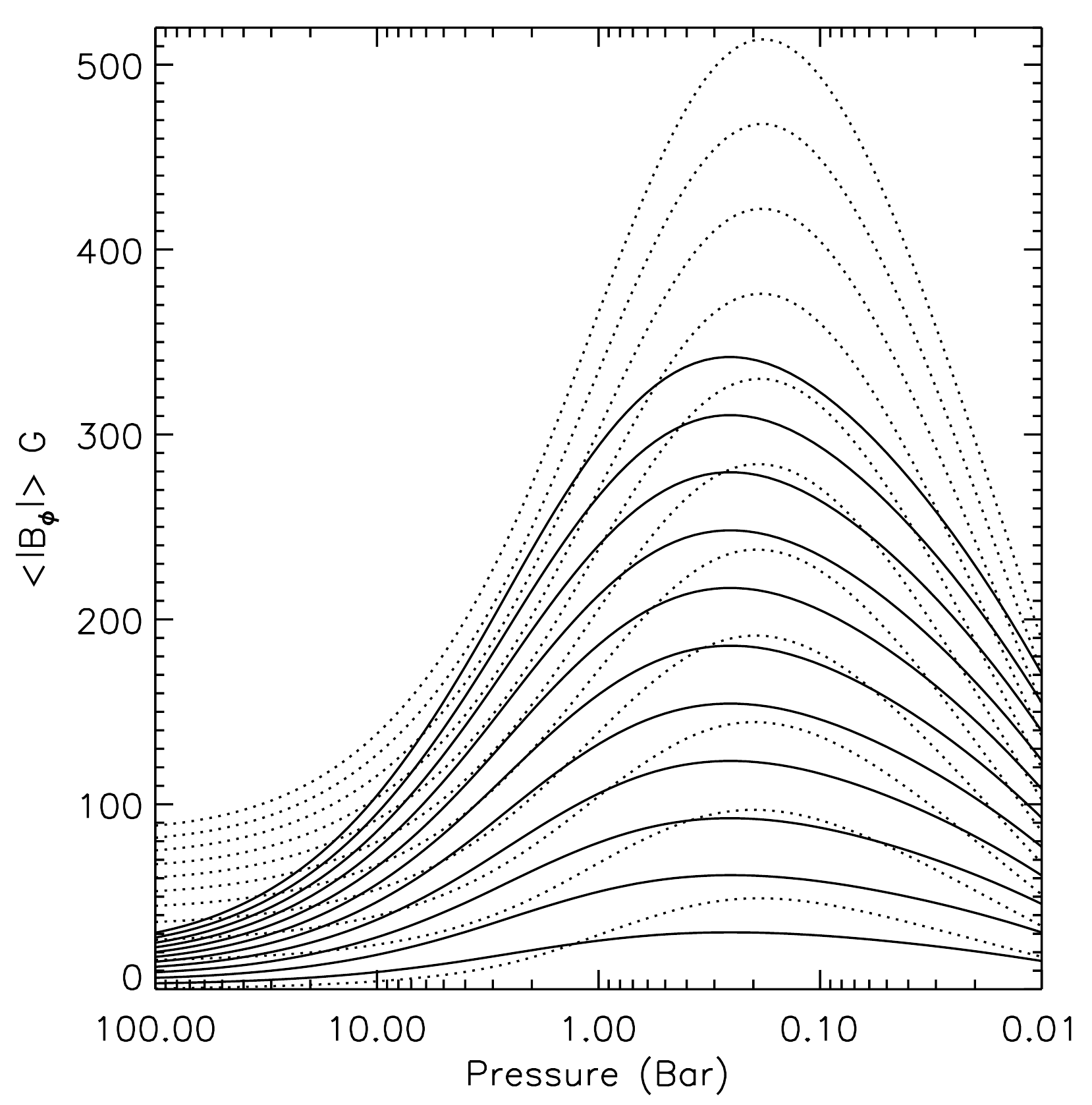}
 \caption{Time evolution of horizontally averaged toroidal field as a function of depth for models M7b1 (solid lines) and M7b2 (dotted lines). Induced toroidal field peaks around 0.1-0.2 Bar, where vertical shear is large. The field strength increases with time starting at 330$P_{\mathrm{rot}}$ and increasing in increments of $\sim$115$P_{\mathrm{rot}}$ to 1500$P_{\mathrm{rot}}$.}
 \label{fig:bphitime}
\end{figure}
In hotter models the induced toroidal field strength grows rapidly and peaks around 0.1-0.2 Bar, as seen in Figure~\ref{fig:bphitime}, where zonal wind shear is strong (see Figure~\ref{fig:zmzw-rad}). One can see that the peak field amplitude grows in time and that the magnetic layer spreads. With maximum wind speeds of $\sim$km/s, the equipartition field strength is of order 100-200G depending on the depth of the flow (density). In the hot models, toroidal field strengths easily reach values in excess of this (Figure~\ref{fig:bphitime} and Table 1). Stated another way, the Alfven speed ($v_A=B/\sqrt{4\pi\rho}$) is as large, or larger than, the peak wind speed.\footnote{The magnetic pressure at these field strengths is equivalent to $\sim$mBar} 

The stability of the field to buoyancy instability is determined not only by the field strength, but by its radial structure, the stabilizing effect of gravity (the Brunt-Vaisala frequency, N) and the diffusion coefficients (thermal, viscous and magnetic) \citep{sv09}. In these models the Richardson number, N$^{2}$/(dU/dz)$^{2}$, is approximately $\sim$10 so winds are hydrodynamically stable. For hot, magnetic models the ideal (non-diffusive) instability criterion outlined in \cite{acheson78} is satisfied at some times and in some locations and becomes more frequent the hotter the model. However, diffusive effects are likely very important in these models. For example, \cite{sv09} showed that if $\eta < \kappa$ then buoyancy instability is enhanced by double diffusive effects. Models just slightly hotter than those presented would have $\eta <\kappa$ and we note that we were not able to run such models with magnetic fields (hotter hydrodynamic models are possible) as an instability (of unknown nature) always developed. Further, the short timescales associated with Newtonian cooling at the surface are physically equivalent to much larger values of $\kappa$. For example, a cooling time of 10$^{4}$s is something like a $\kappa$ of 10$^{14}\mathrm{cm^{2}/s}$, resulting in $\eta \ll \kappa$, again, possibly enhancing buoyancy instability in hot models. The resolution in these models is insufficient to capture a buoyancy instability if one were to develop. However, given the arguments above, a buoyancy instability appears probable in the hottest atmospheres. Turbulence arising from such an instability may provide a downward flux of heat which could contribute to planetary radius inflation, as suggested by \cite{youdin10}. Therefore, more work should be done investigating buoyancy instability under these hot Jupiter conditions. 

\begin{figure}
\centering
\includegraphics[width=3.5in]{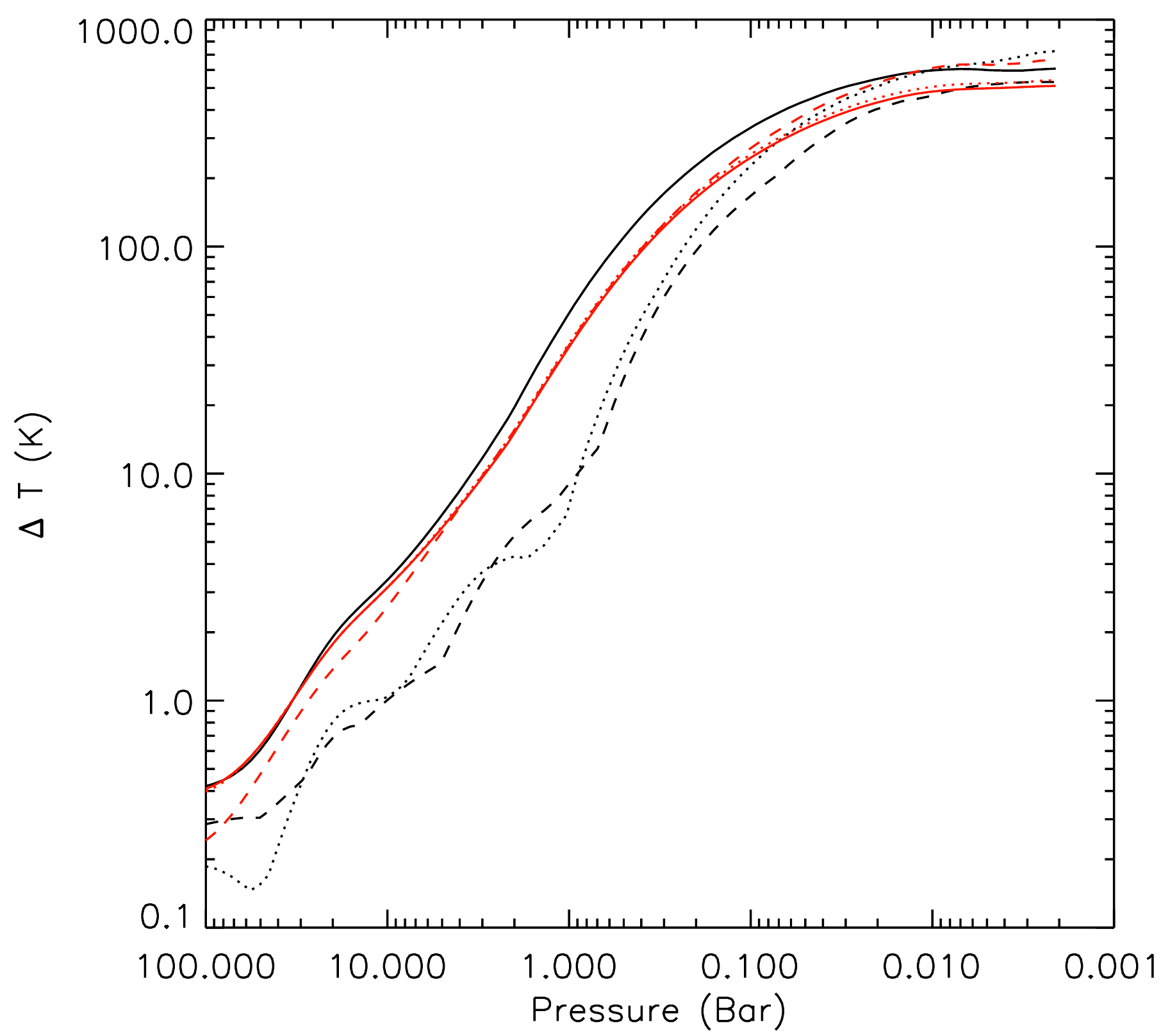}
\caption{Day-Night temperature differential as a function of pressure. Solid line is the hydrodynamic version, dotted lines are 10G MHD models and dashed lines are 30G MHD models. Red lines represent cool models, M2, M2b1 and M2b2, and black lines represent hot models, M7, M7b1 and M7b2.}
\label{fig:dntdiff}
\end{figure}

Because the magnetic diffusivity is not a function of all space we are not able to investigate the instability proposed by \cite{menou12b}. However, it is possible that differential Ohmic heating affects dynamics. Ohmic heating causes the temperature to rise throughout the atmosphere. But because the current is strongest on the nightside, Ohmic heating is stronger there, causing the nightside to heat up more than the dayside thereby reducing day-night temperature differences, and hence, reducing atmospheric forcing (see Figure~\ref{fig:dntdiff}). In the models presented here, this effect is relatively small and may not persist if the magnetic diffusivity were a function of all space. In that case, atmospheric flow-field coupling may be reduced on the nightside which could lower the current (and hence Ohmic heating). On the other hand, the magnetic diffusivity would be locally higher which would increase Ohmic heating and it is unclear how these two effects would interact. We will be addressing this in future work. 
\subsection{Magnetic Scaling Laws}
\begin{figure}
\centering
\includegraphics[width=3.5in]{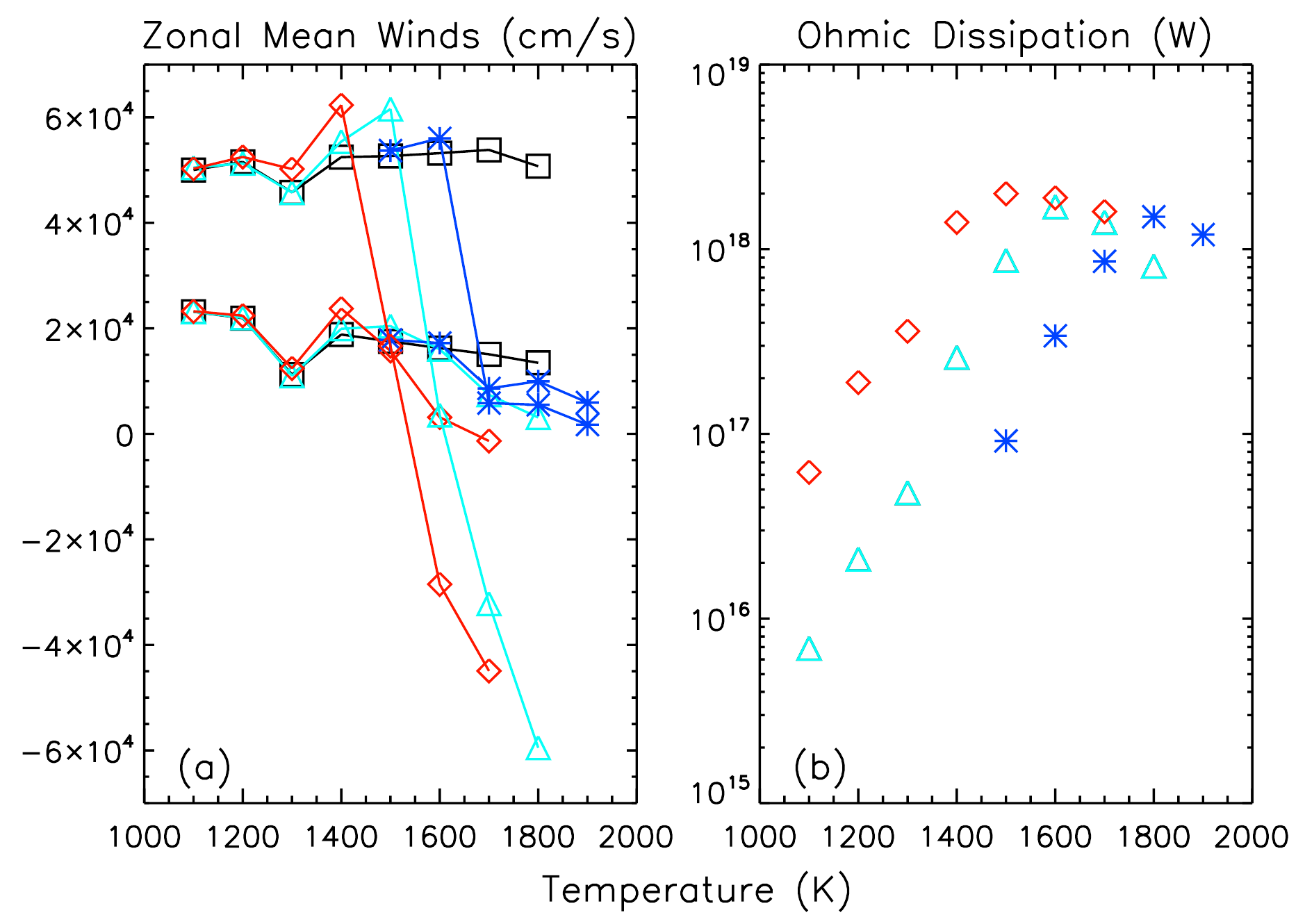}
\caption{Zonal-mean zonal winds at two heights (10mBar, higher speeds and 2 Bar, lower speeds), averaged near the equator as a function of temperature (a) black squares represent hydrodynamic models, cyan triangles represent 10G field, red diamonds represent 30G field and blue asterisks represent 3G field strength. Ohmic dissipation (Watts) at 10Bar as a function of temperature is shown with the same symbols (b). }
\label{fig:zmzw+od}
\end{figure}

\cite{menou12} developed scaling laws for magnetic effects coupling both the effects of wind drag and Ohmic dissipation. He showed that as the temperature of an
atmosphere is increased, field-flow coupling is enhanced and a
magnetic field can more readily slow wind
speeds, which in turn, will reduce Ohmic dissipation. His theory
predicted a temperature at which Ohmic dissipation peaks and
furthermore, predicted an anticorrelation between displacement of the
hot-spot on a planetary surface and the radius anomaly. 
With our suite of models we can test those predictions in the context of this more complete model. In Figure~\ref{fig:zmzw+od} we show zonal mean zonal wind speeds as a function of temperature (a)
at two different pressure levels (2Bar and 10mBar) as well as Ohmic dissipation at 10Bar (b). 
The most obvious result is that the overall behavior produced in our
self-consistent MHD models is remarkably similar to that predicted by
\cite{menou12} (see his Figures 1 and 3): magnetic effects have little
effect on slowing cooler atmospheres but wind speeds drop precipitously
above 1400-1500K and Ohmic dissipation drops above 1500-1600K with 
larger field strengths peaking at lower temperatures.
There are, however, a few interesting and important differences. 
First, with respect to Ohmic
dissipation, the amplitude of the Ohmic heating is 2-3 orders of
magnitude lower than that predicted by previous authors
\citep{bat10,bat11,menou12}. Second, with regard to the magnetic effects on winds, hotter models
show zonal mean zonal winds which {\it reverse} sign
and become retrograde (or westward). That is, the effect of magnetic
fields is not simply to slow the flow, but can act to change the
overall structure of the flow (see Figure~\ref{fig:comp-tempwinds-700}).  These differences will be addressed in
the next sections.
\subsection{Ohmic Dissipation}
There are several reasons the Ohmic dissipation is lower in these 
self-consistent simulations than previously predicted, most are 
physical limitations of the reduced equations employed in previous
attempts to include magnetic effects \citep{bat10,perna10a,perna10b,rm13} and one 
may be a limitation of our anelastic model. 

The first step in estimating ohmic dissipation in many previous models is estimating 
the current using a reduced form of Ohm's Law,
\begin{eqnarray}
{\bf J}=\frac{1}{4\pi\eta}\left(\left( \vvec \times \Bvec \right)+\Evec \right)\approx \frac{1}{4\pi\eta}\left(\vvec \times \Bvec \right) \mathrm{,}
\end{eqnarray}
neglecting the electric field. Retaining only the latitudinal current and
assuming the the radial velocity is sufficiently small, the current is approximated as:
\begin{eqnarray}
J_{\theta}=\frac{1}{4\pi\eta}\left(v_{r}B_{\phi}-v_{\phi}B_{r}\right)\approx -\frac{v_{\phi}B_{r}}{4\pi\eta} \mathrm{.}
\end{eqnarray}
Using this current the Ohmic dissipation and Lorentz Force can be estimated:
\begin{eqnarray}
P=4\pi\eta {\bf J}^{2}\approx \frac {v_{\phi}^{2} B_{r}^{2}}{4\pi\eta} \mathrm{,}\\
\Jvec \times \Bvec\approx -\frac{v_{\phi}B_{r}^{2}}{4\pi\eta} \mathrm{.}
\end{eqnarray}

\begin{figure}
\centering
\includegraphics[width=3.5in]{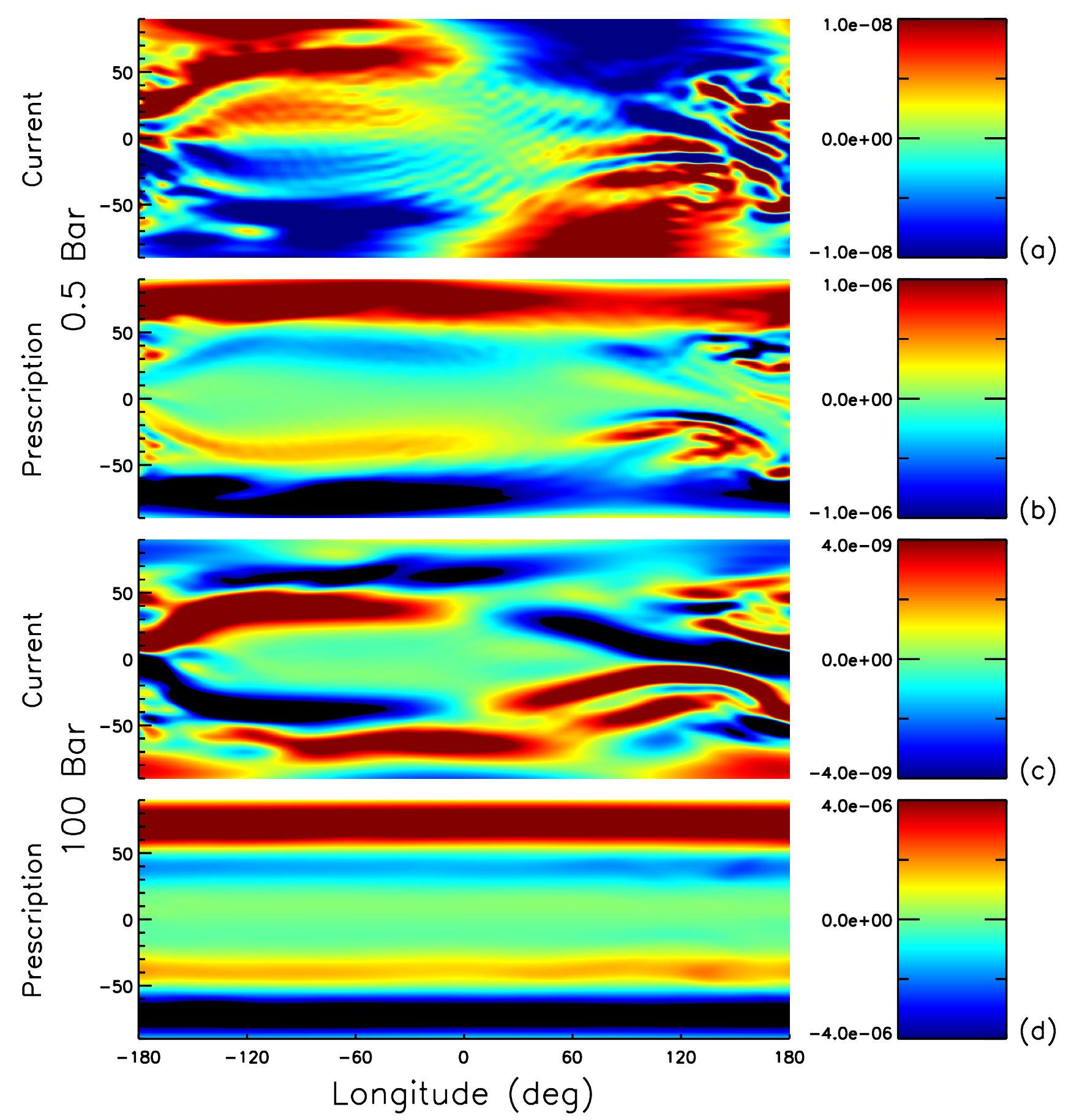}
\caption{Comparison of current calculated using the approximation encapsulated in Equations (13-15) (b,d, labeled ``Prescription'') and current calculated from Ampere's Law (a,c, labeled ``Current'') at 0.5 Bar (a,b) and at 100 Bar (c,d), units are Gauss/cm. One can see that the spatial distribution is not well represented by the approximation and, more importantly, the amplitude recovered using the approximation is two orders of magnitude larger than the actual current.}
\label{fig:compcur}
\end{figure}

\begin{figure}
\centering
\includegraphics[width=3.5in]{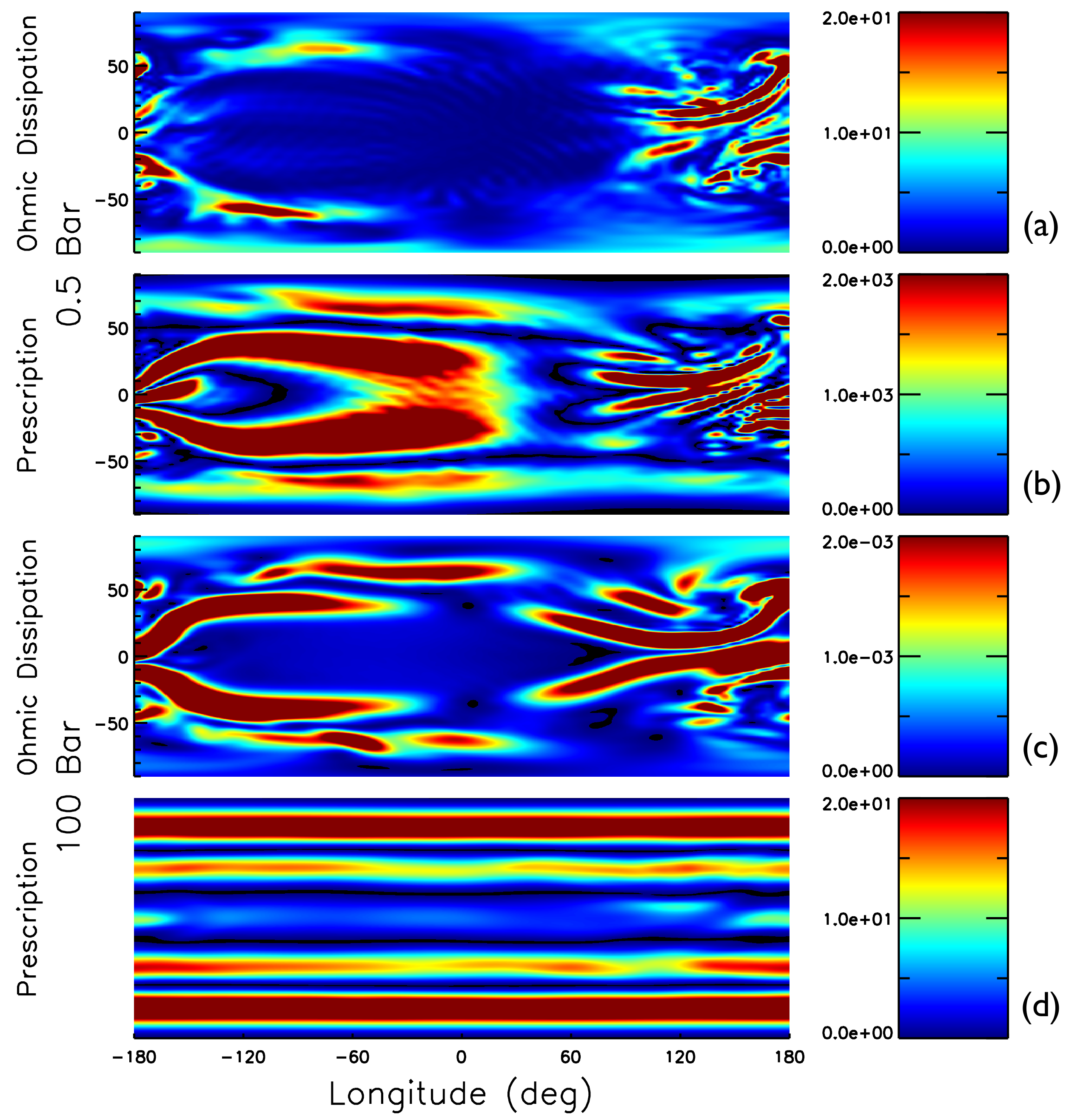}
\caption{Comparison of Ohmic dissipation calculated using Equation (14) (b,d, labeled ``Prescription'') compared to that calculated in simulation (a,c, labeled ``Ohmic Dissipation''), units are Watts/$\mathrm{cm}^2$. In Equation (14) we use velocities from this simulation and a constant magnetic field strength of 30G. }
\label{fig:compohmd}
\end{figure}

The validity of the first step in this estimate, calculating the current, can be tested by comparing Equation (13) with the latitudinal current calculated from Ampere's Law $\Jvec=\curl \Bvec/\mu$, this comparison is shown in Figure~\ref{fig:compcur}. For the approximation, we use velocities taken from this simulation and a constant magnetic field strength, as has been done previously \citep{perna10a,rm13}. One can see there that the approximation overestimates the peak current (note amplitudes on colorbar) by two orders of magnitude and does not accurately reflect the spatial distribution of the current. This overestimate carries over to the Ohmic dissipation, where the approximation (shown in Figure~\ref{fig:compohmd}b,d) overestimates the peak Ohmic dissipation (a,c) by 2-4 orders of magnitude and again, fails to account for the spatial distribution. Upon integration in radius this leads to Ohmic dissipation rates which are generally 2-3 orders of magnitude smaller than estimates using the prescription outlined in Equations (12)-(15) above (values at 10 Bar can be seen in Figure~\ref{fig:zmzw+od} and integrated values can be seen in Table 1). 
The inconsistency between our models and previous estimates stems from neglecting the electric field in  Ohms law, which is only valid in the limit that $\partial{{\bf B}}/\partial{t}$ is small.  This is also equivalent to assuming the low magnetic Reynolds number limit ($R_{m} \lesssim 1$).  With horizontal flow speeds of order $10^{5}$cm/s, length scales of order $10^{9}$cm and magnetic diffusivities as seen in Figure~\ref{fig:tpeta}b, our hottest models have $R_{m}$ 100-1000, clearly in violation of the approximation.  One can also see in Figures~\ref{fig:fieldevolution-back} and \ref{fig:bphitime} that magnetic field evolution is substantial, and neglecting this evolution leads to large discrepancies between our self-consistent model and previous estimates.

Considering a rather hot model of HD209458b with an interior temperature of 2500K leads to $\sim 10^{18}$ W of Ohmic heating at 10Bar. There are many estimates for the amount of heating required to inflate the radius. \cite{show02} state 10\% of stellar insolation is required at 5 Bar, 1\% at 100Bar and 0.08\% if the energy is deposited at the center. \cite{bat11} find that 1-3\% of stellar insolation integrated throughout the planet is required. For a stellar insolation of $3.3 \times 10^{22}$W, we find an efficiency of conversion from incident stellar flux to Ohmic dissipation (i.e. the ratio of integrated Ohmic dissipation to incident stellar flux) of 0.006\% for the most favorable case with $2.2 \times 10^{19}$ W of dissipation integrated below 10 Bar, see Table 1. \cite{bat10} require $4 \times 10^{18}$W in the convective interior to inflate the radius for a solar metallicity, no core model of HD209458b. For HD209458b 1\% of stellar insolation would require $\sim 10^{20}$ W, so our peak atmospheric heating of $\sim 10^{18}$ W would appear too weak to explain the inflated radius by two orders of magnitude. We can extrapolate our heating in the wind zone of the planet to interior regions as in \cite{rm13} by using the scalings of \cite{wu13}, assuming conservation of currents between the atmosphere and interior. To order of magnitude precision, \cite{wu13} predict that the Ohmic dissipation in the interior (below the bottom boundary of our model) is a factor of $\sim z_{\mathrm{wind}}/R_p$ lower than that in the atmosphere. The thickness of our modeled atmosphere, from 2 mbar to 200 bar, is $\sim 1.2 \times 10^9$cm (about 14\% of the stellar radius). Hence, given an integrated atmospheric Ohmic heating of $1 \times 10^{18}$ W, we expect a total Ohmic heating rate below the bottom boundary of $1.4 \times 10^{17}$W. This estimate is more than an order of magnitude below the necessary $4 \times 10^{18}$W interior dissipation from \cite{bat10}. More realistic models for HD209458b, in terms of temperature, appear too weak by up to three orders of magnitude (1300K surface temperature, 2000K interior temperature, 10G field). The effects of various parameters, and the possibility that our lower wind speeds contribute to this shortcoming, will be discussed in Section 3.5.
\subsection{Zonal Winds}
Previous work implementing the Lorentz force in GCMs of the atmospheres of hot Jupiters (e.g. \cite{perna10a,rm13}) did so by incorporating a Rayleigh drag, similar to the implementation of  \cite{Schneider09} previously for Jupiter itself. This kinematic method slows the zonal wind through adding a drag term $= -u/\tau_{\mathrm{mag}}$ to the zonal component momentum equation, where the magnetic timescale $\tau_{mag}$ (or the inverse relaxation coefficient for Rayleigh drag) is defined as $\tau_{mag}=4\pi\rho\eta/B^{2}$ \citep{rm13}.  
For Jupiter \cite{Schneider09} find that a reduced heat flux combined with strong drag can inhibit superrotation and introduce retrograde flow (subrotation).  However, no previous work on hot Jupiter atmospheres has found this effect.  Similar to the discussion above, in treating the Lorentz force simply as a drag on the zonal component of the momentum equation, one must neglect the evolution of the magnetic field because the Lorentz force only acts like a drag on the fluid velocity perpendicular to the magnetic field.  Therefore, one must assume a pure dipole in order to assume the zonal flow is continuously perpendicular to it.  As can be seen in Figure~\ref{fig:fieldevolution-back}, this is not a good assumption, particularly for hot models (high $R_{m}$).
\begin{figure}
\centering
\includegraphics[width=3.5in]{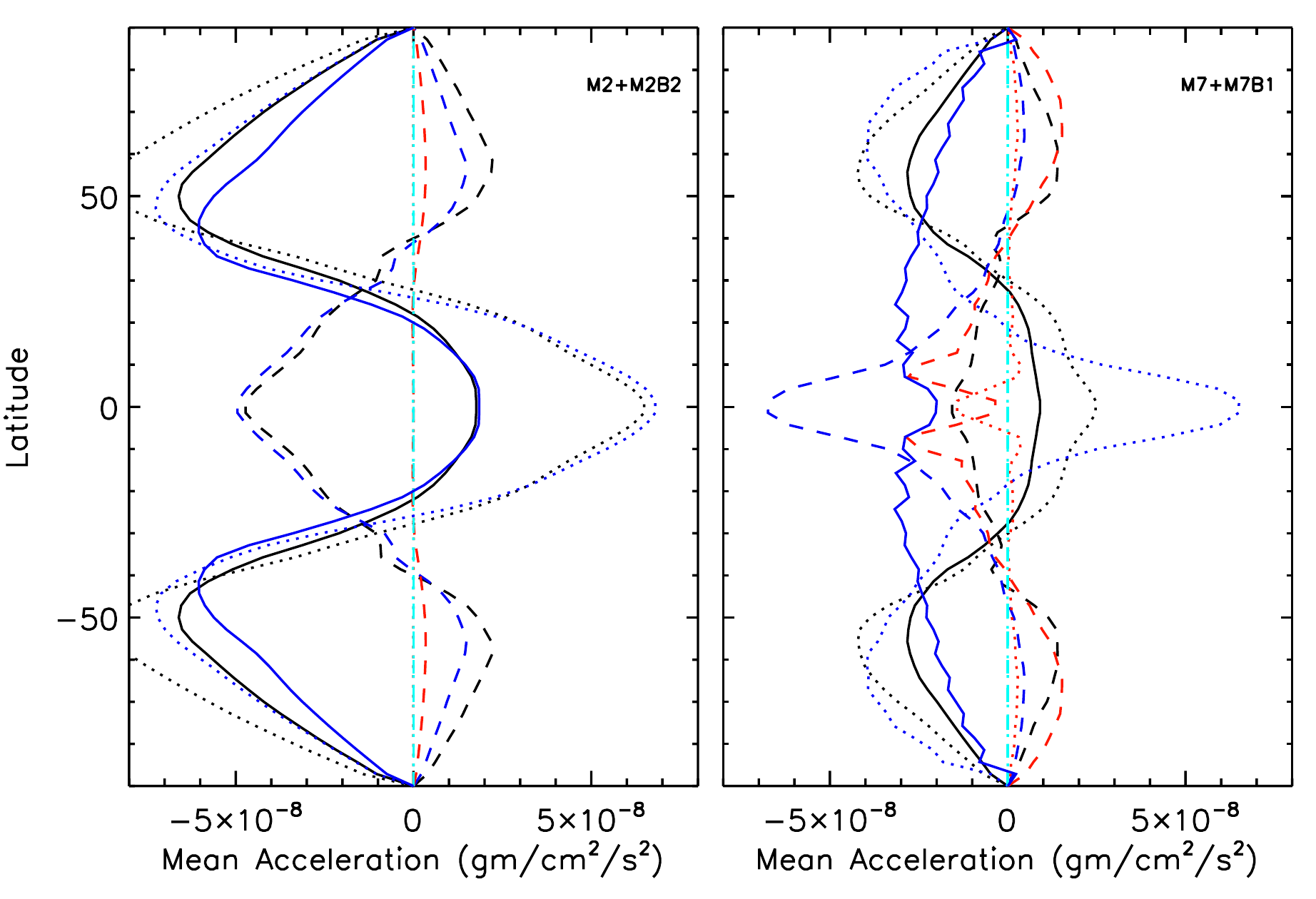}
\caption{Left: Cool model M2 (hydrodynamic model) and M2B2 (MHD model). Right: Hot model M7 (hydrodynamic model) and M7B1 (MHD model). Lines represent components of Reynolds and Maxwell stresses. Dotted and dashed lines represent horizontal and vertical components of Reynolds stresses, respectively, black lines represent the purely hydrodynamic models, while blue lines represent the same stresses in the MHD models. Red lines represent Maxwell stresses in the MHD models and solid lines represent the sum of the stresses with black representing the hydrodynamic model and blue representing the MHD model. Cyan lines represent viscous stresses. In hot models, magnetic stresses are negative (westward) near the equator and therefore, act to reverse wind speeds.} 
\label{fig:stresses}
\end{figure}
\begin{figure*}
\centering
\includegraphics[width=5in]{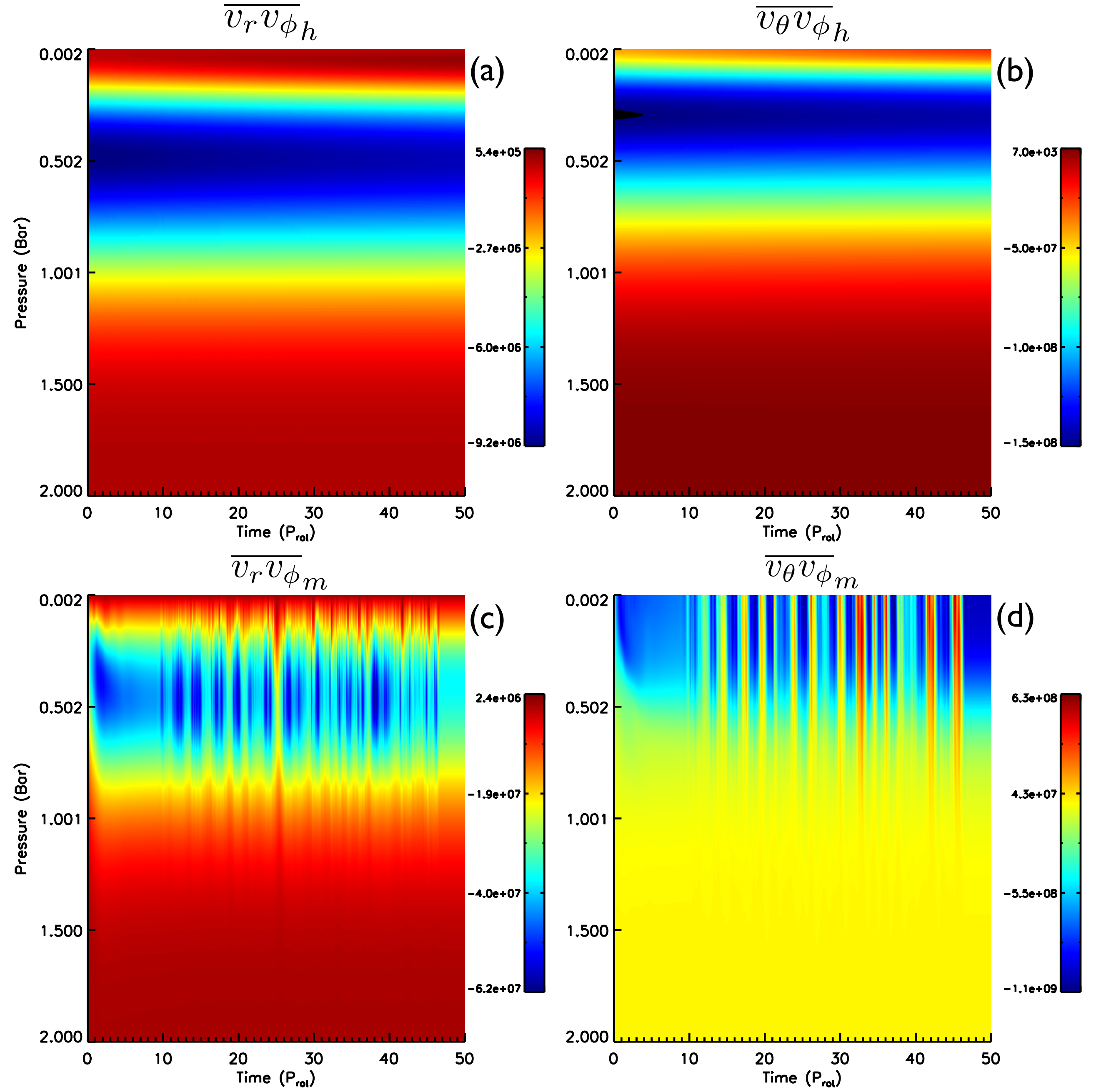}
\caption{Reynolds stresses contributing to zonal flow in hydrodynamic models (a,b, Model M7) and MHD models (c,d, Model M7b1) at the equator as a function of time and pressure in the atmosphere.  One can see that the velocity fluctuations are anti-correlated and steady in time in the hydrodynamic models but this correlation is intermittently disrupted with opposite correlations in the MHD cases.  This results in variability of the equatorial zonal wind.}
\label{fig:correlations}
\end{figure*}
\begin{figure}
\centering
\includegraphics[width=3.5in]{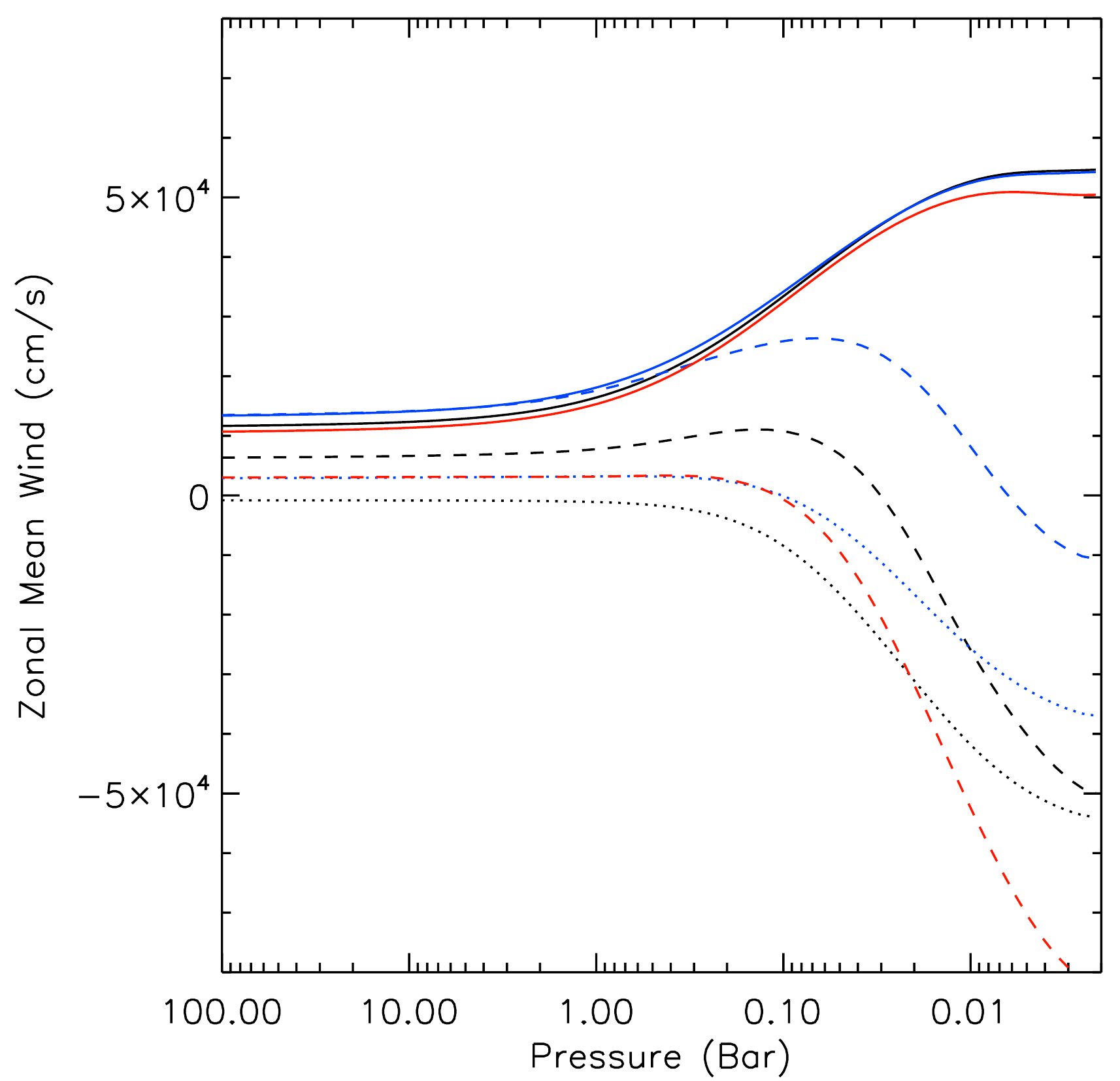}
\caption{Zonal mean zonal wind, averaged over low latitudes, as a function of pressure. Black lines represent models M7, blue lines represent M8 and red lines represent M6, with solid lines hydrodynamic models, dashed 10G models and dotted 30G models.}
\label{fig:zmzw-rad}
\end{figure}
\begin{figure}
\centering
\includegraphics[width=3.5in]{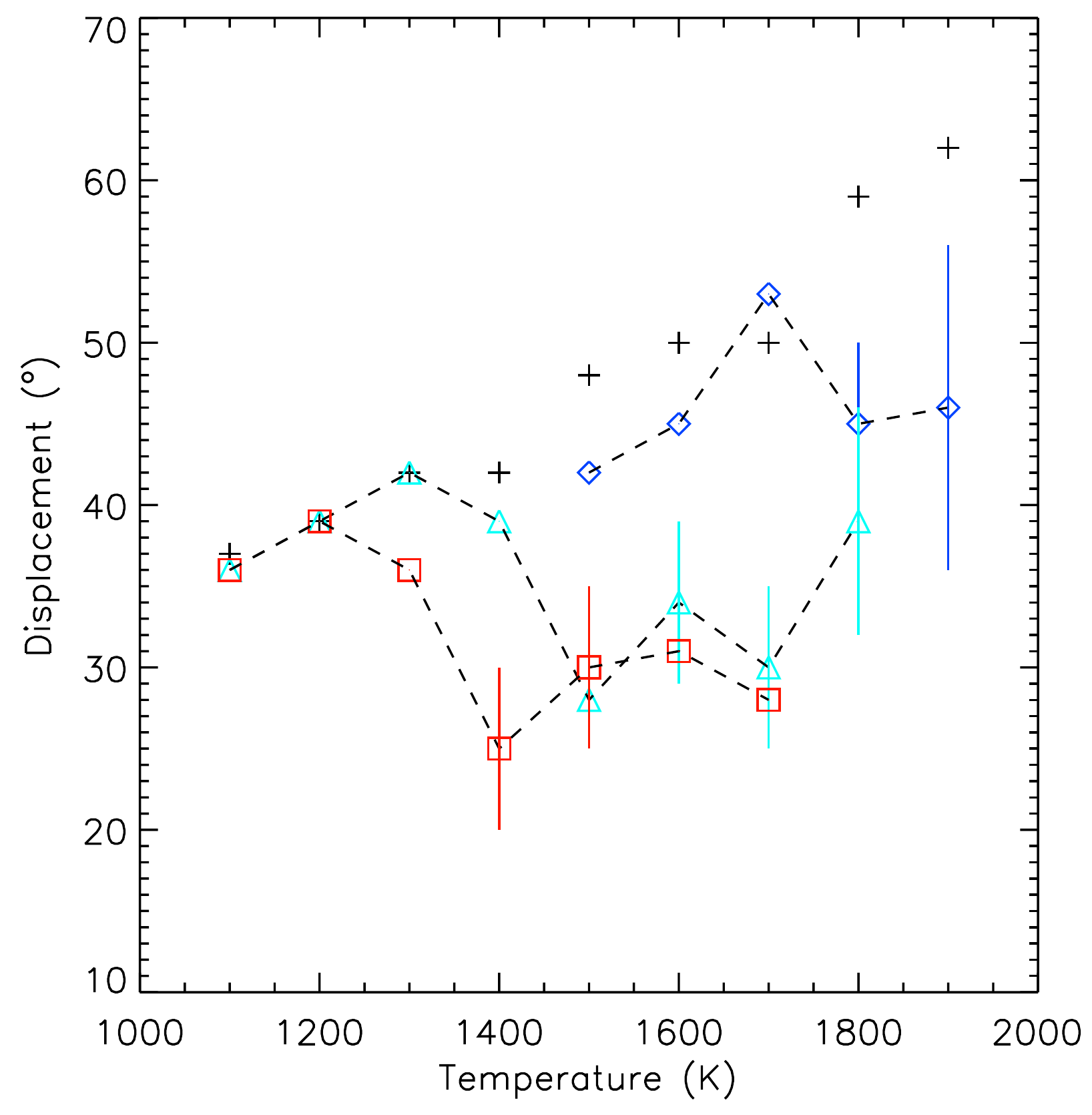}
\caption{Hotspot displacement as a function of temperature. Black pluses denote hydrodynamic models, blue diamonds represent 3G magnetic models, cyan triangles represent 10G models and red squares represent 30G models. Vertical lines represent the range over which the hotspot displacement varies in 100 P$_{\mathrm{rot}}$.}
\label{fig:hotspot}
\end{figure}
Here, we find that the effect of magnetic fields on zonal winds can be treated as drag only in a narrow range of temperatures, where $R_{m}$ is low.  At higher temperatures (higher $R_{m}$), the effect of magnetic fields is not simply to act as drag.  At the lowest temperatures the magnetic field has virtually no effect on the wind speeds or geometry, the magnetic diffusivity is sufficiently high that magnetic the flow just slips past the magnetic field unimpeded.  At 1400-1600K, depending on the field strength, the magnetic field starts to slow the wind in the upper and lower atmosphere.  For higher temperatures, and at larger field strengths, the zonal mean zonal wind reverses direction, becoming westward. The magnetic timescale, defined as $\tau_{mag}=4\pi\rho\eta/B^{2}$, varies from $\sim$ 10$^{4}$s-10$^{7}$s in our models (see Table 1). Near the surface where the radiative timescale is $10^{4}$s, the magnetic timescale can become shorter than the radiative timescale. When this occurs, magnetic effects begin to dominate the surface flows. As the magnetic timescale is reduced magnetic effects change from: no effects ($\tau_{mag} > \tau_{rad}$), to oscillatory mean flows ($\tau_{mag} \sim \tau_{rad}$) to reversed mean flows ($\tau_{mag} < \tau_{rad}$).\footnote{We use the radiative timescale here because it is fixed as opposed to the strongly (time and location) varying dynamical timescale. Furthermore, the radiative timescale does mimic, at some level, the wave forcing that contributes to superrotation.}

We can investigate this behavior by looking at the mean azimuthal momentum equation: 
\begin{eqnarray}
\dxdy{\overline {\rho v_{\phi}}}{t}&\approx&-\nabla\cdot\left(\rhobar\overline{\vvec\vvec}\right)+\frac{\nabla\cdot\left(\overline{\Bvec\Bvec}\right)}{4\pi}\\
&\approx&-\frac{1}{r^{2}}\dxdy{\left(r^{2}\overline{\rho v_{r}v_{\phi}}\right)}{r}-\frac{1}{r\sin\theta}\dxdy{\left(\overline {\rho v_{\theta}v_{\phi}}\sin\theta\right)}{\theta}\\
&+&\frac{1}{r^{2}}\dxdy{\left(r^{2}\overline{B_{r}B_{\phi}}\right)}{r}+\frac{1}{r\sin\theta}\dxdy{ \left(\overline{B_{\theta}B_{\phi}}\right)}{\theta} \nonumber
\end{eqnarray} 
where $\theta$ is the colatitude and we have neglected the Coriolis and viscous terms as small and pressure gradient terms (magnetic and gas) are zero on horizontal average. In a purely hydrodynamic model, the superrotation is driven by a slight imbalance between the vertical flux of eddy momentum (first term on RHS of Equation (17), which drives retrograde flows) and the horizontal flux of eddy momentum (second term on RHS of (17), which drives prograde flows) as described in \cite{showpolv11}. We find the same imbalance in our hydrodynamic and cool MHD models, see Figure~\ref{fig:stresses}a. That figure shows those stresses integrated from 20-2mBar, early in the simulation, before they have reached a steady state (model M7B1 never reaches a steady state and mean winds are time-dependent, see below) in order to better ascertain the causal connection between westward jets and stresses. There we see that the horizontal convergence of eddy flux leads to superrotation at the equator in our cool model for both hydrodynamic (black lines) and MHD models (blue lines). The Maxwell stresses, terms 3 and 4 on the RHS of Equation (17), in this model are negligible (red lines), as are the viscous stresses in all models (cyan lines). The only apparent effect of the magnetic field in this cool model is to slightly narrow the super-rotating equatorial jet. Things are substantially different in a hot model (Figure~\ref{fig:stresses}b). There we again see that the hydrodynamic model (black lines) shows a slight imbalance between vertical and horizontal components of eddy fluxes favoring horizontal eddy fluxes, leading to superrotation at the equator. However, in this case the Maxwell stresses are not negligible and are both negative near the equator (red lines). Therefore, the slight imbalance that favored superrotation at the equator in the hydrodynamic models is flipped in MHD models and westward flows result.

The sense of both components of the Maxwell stress can be understood in the following manner. The induction of the toroidal field is dominated by the radial gradient of the azimuthal flow $B_{\phi,t}\sim B_{r} u_{\phi,r}$. Near the equator in the Northern Hemisphere (NH), $B_{r} > 0$ and $u_{\phi}$ is increasing outward, so that the induced $B_{\phi}$ near the equator in the NH is positive (Figure~\ref{fig:fieldevolution-back}). Since $B_{r}$ is also positive $B_{r}B_{\phi} > 0$, however, $B_{\phi}$ peaks lower in the atmosphere (see Figure~\ref{fig:bphitime}), where $\eta$ is lower and similarly, $B_{r}$ peaks lower in the atmosphere since it is a dipole. Therefore, $B_{r}B_{\phi}$ is decreasing outward, so the third term on the RHS of Equation (17) is negative (leading to westward flow). Similarly, near the equator $B_{\theta}$ is decreasing (becoming more negative) and $B_{\phi}$ is increasing (becoming more positive), so the product is decreasing (becoming more negative) toward the equator, as $\theta$, the colatitude, increases. Therefore, the fourth term on the RHS of Equation (17) is also negative\footnote{Note though that the latitudinal magnetic field is quickly altered by the flows and this becomes hemispherically dependent, and the argument is less simple.}. Hence, when Maxwell stress amplitudes become comparable to the difference between the opposing horizontal and vertical hydrodynamic stresses, magnetism can act to reverse the sign of the zonal-mean flow.\footnote{A reversed initial dipolar field geometry would result in the same westward flow. The induced toroidal field would be negative in the NH, but $B_{r}$ would also be negative, so $B_{r}B_{\phi} > 0$ and the amplitude would still decrease outward. Similarly $B_{\theta}$ would be positive at the equator and the product $B_{\theta}B_{\phi}$ would be negative and decreasing toward the equator.} 

The effect of a magnetic field on zonal winds at high $R_{m}$ has been studied previously by \cite{tdh07}.  Using local two-dimensional (2D) simulations on a $\beta$-plane, they found that the effect of magnetic fields on zonal jets was a subtle one in which magnetic fields corrupted the time coherence between velocity components contributing to zonal jets\footnote{Although those results were in 2D the same result has been found in 3D, D. Hughes private communcation.}.  Here we find something similar, in Figure~\ref{fig:correlations} we show the correlations between velocity components as a function of time and radius in the hydrodynamic (a,b) and MHD (c,d) cases.  There we see that the  magnetic models show periodic disruptions of the velocity correlations which reduces the driving of zonal winds.  The timescale for the disruption of coherence occurs on approximately an Alfven time.  This indicates the typical Rossby-gravity waves which contribute to the eastward zonal jet are additionally influenced by Alfven waves in the high $R_{m}$ limit.

Despite these effects, the reversed and severely dragged mean zonal winds are limited to the upper layers of the atmosphere (pressure levels below 0.1Bar, where the $R_{m}$ is largest), while at deeper levels (where $R_{m}$ is small) the winds are merely dragged, see Figure~\ref{fig:zmzw-rad}. This results in a reduction of the hot spot displacement eastward of the sub-stellar point. Figure~\ref{fig:hotspot} shows the hot spot displacement, at 60mBar (a nominal depth for the photosphere), as a function of temperature and magnetic field strength. Vertical lines on some data points indicate the variability of the hot spot displacement over 100 P$_{rot}$. For hydrodynamic models (shown with pluses), the hot spot displacement increases with temperature from $\sim 35-60^{\circ}$. If one considers reasonable field strengths (3,10G), hot spot displacement is affected significantly only at temperatures larger than 1500K, where it is reduced by as much as 20$^{\circ}$, but can vary by $\sim 10^{\circ}$. Large observed hotspot displacements therefore imply either negligible magnetic field strength or low surface temperatures. High surface temperatures with large hot spot displacements could therefore provide an upper limit on magnetic field strength. Notably, the hot spot displacement has been measured on HD209458b to be $\sim$40$^{\circ}$ (Zellem, private communication), consistent with our hydrodynamic and weak ($\lesssim 3G$) field results which have negligible Ohmic dissipation.\footnote{Consistent with the reversed and oscillating winds higher in the atmosphere, the hotspot varies more wildly aloft, by as much as $30^{\circ}$ and is sometimes displaced westward of the substellar point.}

\subsection{Dependencies}
\begin{figure}
\centering
\includegraphics[width=3.5in]{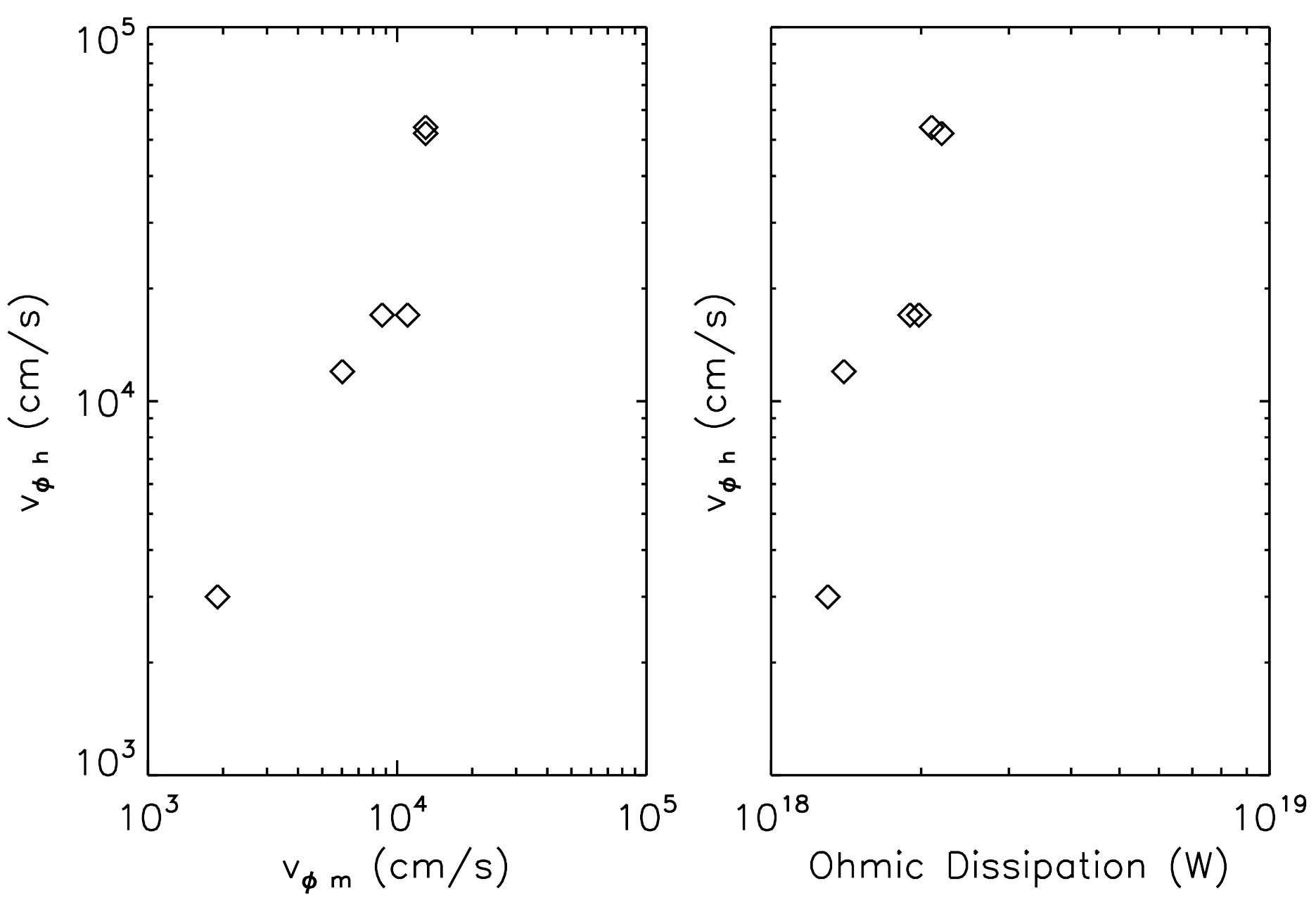}
\caption{(a) Zonal mean zonal wind speeds in hydrodynamic models M7, M7v1, M7v2, M7v3, M7cbc and M7f1 (v$_{\phi h}$) versus zonal mean zonal wind speeds in their MHD counterparts M7b1, M7v1b1, M7v2b1, M7v3b1, M7cbcb1 and M7f1b1 (v$_{\phi m}$). A 20 times increase in hydrodynamic model wind speeds results in only a five fold increase in MHD model wind speeds. (b) Zonal mean zonal wind speeds in the same hydrodynamic models as in (a) versus Ohmic dissipation in those models at 10Bar. A 20 times increase in hydrodynamic model wind speeds results in only a two times increase in Ohmic dissipation.}
\label{fig:vphvpmod}
\end{figure}

Besides temperature and magnetic field strength, we investigated the effect of lowered diffusivities (viscous and thermal, M7v1*, M7v2*, M7v3*), velocity boundary conditions (no slip instead of stress free, M7cbc*) and increased forcing (1200K instead of 800K, M7f1*) all at the same reference temperature (1700K) and field strength (10G). Lowered diffusivities at depth and larger forcing lead to faster wind speeds at depth and increased Ohmic dissipation. No slip boundary conditions lead to reduced wind speeds and hence, lower Ohmic dissipation. Figure~\ref{fig:vphvpmod}a shows the zonal mean zonal wind speeds in hydrodynamic models ($v_{\phi h}$) versus the zonal mean zonal wind speeds in their MHD counterparts ($v_{\phi m}$). There we see a 20x increase in hydrodynamic wind speeds results in only a 5x increase in MHD wind speeds. Furthermore, this 5x increase in MHD wind speed results in only a 2x increase in Ohmic dissipation (Figure~\ref{fig:vphvpmod}b). Therefore, it appears that 1) the simple prescription in which Ohmic dissipation is proportional to zonal wind speed squared is not an accurate representation and 2) increased wind speeds are unlikely to be able to close the (large) gap between the Ohmic dissipation calculated and the values typically considered necessary to explain the inflated radii of massive hot Jupiters. 

\section{Discussion}
We have shown the basic effects of magnetism in the atmospheres of hot Jupiters for a range of temperatures. At low temperatures, ionization is sufficiently low that the flow simply slips past magnetic field lines unaffected. Around 1400-1500K, the behavior changes. Temperatures are large enough to allow sufficient ionization for field-flow coupling and magnetic effects, such as magnetic tension, magnetic pressure and Ohmic dissipation become relevant. This qualitative result has been found previously by \cite{menou12}. 
We find that, even at its peak, Ohmic dissipation is substantially smaller than previously calculated and generally at least an order of magnitude lower than the requisite heating required to explain many hot Jupiter radii. We find that increasing wind speeds, changing boundary conditions, increasing forcing, varying temperature and varying field strength does not substantially alter this picture. However, to adequately evaluate whether Ohmic dissipation is able to inflate individual objects it may be necessary to couple evolutionary models with the heating in a more sophisticated manner. It is likely that Ohmic dissipation may be able to explain the radii of lower mass, inflated planets as found in \cite{huang12}. Furthermore, since the most efficient heating (in terms of radius inflation) should occur deeper in the atmosphere, it may be necessary to better resolve the dynamics of the deep atmosphere, where shallow, hydrostatic models are not optimal. 

With regard to the magnetic effects on hot Jupiter winds and recirculation efficiency we find a more complicated picture than expected. As the magnetic timescale is progressively decreased (atmosphere becomes hotter and/or field becomes stronger) MHD effects progress from pure drag, to oscillatory mean flows, to stationary but westward mean flows. Therefore, high in the atmospheres of hot Jupiters with substantial field strength magnetic effects could cause the winds to vary on short timescales (timescales of order tens of rotation periods) and even cause mean flows and hot spot displacements which are westward. This leads to hot spot displacements which are reduced compared to their hydrodynamic counterparts and which can show substantial variability. 

Additionally, we find that magnetic field evolution is more complicated than the simple kinematic picture. As temperatures rise, magnetic diffusivity drops while induced fields and their vertical gradients grow resulting in more conducive conditions for magnetic buoyancy instability. Although we can not resolve the instability here, numerical instability coupled with favorable conditions indicate such an instability may be possible in hot, magnetic atmospheres.  

The models presented here have several shortcomings which should be kept in mind when considering their results. First, these models are anelastic and hence, preclude at some level, the fast wind speeds commonly recovered in typical models of hot Jupiter atmospheres. While this may have an effect on the degree of Ohmic heating, the winds at depths where Ohmic dissipation is likely to be the most efficacious are highly uncertain. It is unclear at what level the dynamics of that region are driven by the stellar insolation-driven winds aloft or the convective motions below and more sophisticated models coupling those regions may be necessary to fully understand the dynamics in that region and hence, the Ohmic dissipation there. Second, these models do not account for a fully time and temperature dependent magnetic diffusivity (which should be a function of all space). While we doubt that this physics will have a substantial effect on Ohmic dissipation, it likely will affect the wind structure, particularly high in the atmosphere and it may affect the presence of instabilities which drive turbulence. We are working on including this effect, but we note that the large temperature variation ($\sim$ 1000K) seen in some planets are likely not computationally feasible and similarly, extremely hot planets will not be accessible without substantially increased computational time and/or some extrapolation. Finally, our reference state pressure-temperature profiles are rather crude and more realistic profiles should be used in follow up studies. This will be incorporated into future work looking at individual planets. 

With these caveats in mind, we have three main conclusions:
\begin{enumerate}
\item Ohmic dissiption appears insufficient to explain all of the inflated radii of observed hot Jupiters. 
\item Magnetic effects do not act simply to slow winds, but can have much more complicated, time-dependent effects, which may be observable in phase curves (hot spot displacement). 
\item It appears probable, particularly in hot models, that a magnetic buoyancy instability could proceed, possibly producing turbulence in the atmosphere. 
\end{enumerate}
\section*{Acknowledgments}
We are grateful to A. Cumming, G. Glatzmaier, D. Lin, A. Showman and G. Vasil for helpful discussions. Support for this research was provided by NASA grant NNG06GD44G. Computing was completed on Pleiades at NASA Ames. 
\bibliographystyle{apj} 
\bibliography{hjm} 
\end{document}